\documentclass[preprint, 12pt]{elsarticle}
\biboptions{authoryear}

\usepackage{graphicx} 
\usepackage{animate}
\usepackage{amssymb} 
\usepackage{amsmath}
\usepackage{multirow} 
\usepackage{lineno} 
\usepackage{amsfonts,amsthm,bm}
\allowdisplaybreaks

\begin{document}

\begin{frontmatter}

\journal{Spatial Statistics} 
\title{A spatially varying change points model for monitoring glaucoma progression using visual field data}
\author[1]{Samuel I. Berchuck} 
\author[2]{Jean-Claude Mwanza} 
\author[3]{Joshua L. Warren\corref{cor1}} \ead{joshua.warren@yale.edu}
\cortext[cor1]{Correspondence to: Department of Biostatistics, Yale University, 60 College Street, Ste 213, New Haven, CT, USA 06510}
\address[1]{Department of Statistical Science and Forge, Duke University, NC, USA}
\address[2]{Department of Ophthalmology,  University of North Carolina-Chapel Hill, NC, USA}
\address[3]{Department of Biostatistics, Yale University, New Haven, CT, USA}

\begin{abstract}
Glaucoma disease progression, as measured by visual field (VF) data, is often defined by periods of relative stability followed by an abrupt decrease in visual ability at some point in time. Determining the transition point of the disease trajectory to a more severe state is important clinically for disease management and for avoiding irreversible vision loss. Based on this, we present a unified statistical modeling framework that permits prediction of the timing and spatial location of future vision loss and informs clinical decisions regarding disease progression. The developed method incorporates anatomical information to create a biologically plausible data-generating model. We accomplish this by introducing a spatially varying coefficients model that includes spatially varying change points to detect structural shifts in both the mean and variance process of VF data across both space and time. The VF location-specific change point represents the underlying, and potentially censored, timing of true change in disease trajectory while a multivariate spatial boundary detection structure is introduced that accounts for the complex spatial connectivity of the VF and optic disc. We show that our method improves estimation and prediction of multiple aspects of disease management in comparison to existing methods through simulation and real data application. The R package \texttt{spCP} implements the new methodology. 
\end{abstract}


\end{frontmatter}



\section{Introduction \label{sec:intro}}

Glaucoma is an optic neuropathy that is the leading cause of irreversible vision loss worldwide \citep{tham2014global}. Once a patient is diagnosed, they are at risk of disease progression and are closely monitored by clinicians who must balance the potential of further irreversible deterioration of vision with the costs (financial and physical) of medical and surgical interventions. The visual ability of a glaucoma patient is most often quantified through visual field (VF) examinations, a psychophysical procedure that assesses a patient's field of vision. VF tests are performed during clinical visits over time, producing a time series of spatial referenced data, and the results are monitored for signs of glaucomatous VF progression. Determining if the disease is progressing remains one of the most challenging aspects of glaucoma management, as it is difficult to differentiate true progression from testing variability \citep{vianna2015detect}.

Although VF testing is a common technique for assessing glaucoma progression, there is currently no reference standard approach for converting its output into a treatment decision. The majority of past studies can be grouped into two categories, global- and trend-based analyses. Global methods provide a single summary of the VF data at each visit, thus aggregating and potentially ignoring key information across VF locations \citep{heijl1987package}. These global summaries are then monitored over time to investigate the potential of disease progression. Trend methods analyze the individual time series of sensitivities at each VF location \citep{fitzke1996analysis}. These methods often resort to separately modeling the sensitivities over time assuming a linear relationship, often through use of simple linear regression \citep{katz1997estimating}. The inference obtained from these methods can be a poor diagnostic of progression or have inferior prediction accuracy, in comparison to non-linear models \citep{jampel2011assessment}, and suffers from low power due to ignoring the spatial nature of VF data or aggregating over space \citep{vianna2015detect}. 

In this paper, we develop an innovative method that utilizes spatially varying change points (CPs) to accurately model the VF sensitivities over time. Patients diagnosed with glaucoma are often monitored for years with only slow changes in visual functionality \citep{heijl2002reduction}. It is not until disease progression that notable vision loss occurs, and the deterioration is often swift \citep{jay1993rate}. This disease course inspires a modeling framework that can identify a point of functional change in the course of follow-up. By utilizing the CP framework, the concept of disease progression becomes intrinsically parameterized into the model. Not only does the specification increase flexibility in modeling the trajectory of the VF time series by relaxing the linearity assumption, but the CP represents the point of functional change and can be used to make inferential statements about disease progression. This framework is particularly rich, however, because in addition to there being a functional change in the VF sensitivities upon the start of progression, there is also an increase in testing variance in areas of decreased sensitivity. This inverse variance relationship upon progression is well known in the glaucoma literature, although is often ignored in model development \citep{russell2012relationship}. We use a CP to segment the variance into a stable process that has the flexibility to increase after the point of progression. Finally, we focus on the joint modeling of all available VF data for a patient, accounting for the complex underlying spatial correlation, thereby increasing power to make improved predictions and treatment decisions.

The VF exhibits a complex spatial structure that is best modeled considering the anatomy of the eye. Naive spatial/neighborhood structures that assume neighboring VF locations share information similarly ignore this anatomy. Incorporating this information can aide in properly modeling this spatial dependency and in uncovering important features of disease progression across time. \citet{berchuck2018} introduced a framework for inducing localized neighbors as a function of ocular anatomy using a spatiotemporal boundary detection method. Their method utilized a measure developed by \citet{garway2000mapping} that represents an estimate of the angle that each VF test location's underlying retinal ganglion cells (RGC) enter the optic disc. The Garway-Heath angles are used as a dissimilarity metric, a covariate that explains localized spatial structure. In the present paper, the methodology from \citet{berchuck2018} is extended to the multivariate setting, such that the Garway-Heath angles dictate the neighborhood structure on the VF for multiple spatial parameters, including the spatially varying coefficients and CPs. To our knowledge, this is the first extension of boundary detection using a dissimilarity metric to the multivariate setting.

Our newly developed spatial CPs model with multivariate boundary detection structure has the ability to improve prediction of future VF sensitivities and assess progression for a glaucoma patient. Having the ability to accurately predict future deterioration of the VF will improve a clinician's ability to make timely, data-driven treatment adjustments to preserve vision. Furthermore, a latent CP process is introduced that allows us to spatially interpolate CP values at VF locations where visual damage has yet to occur (i.e., censored locations). This will allow clinicians to determine the location and timing of future vision loss. 

This paper is outlined as follows. Section \ref{sec:data} introduces the VF data while Section \ref{sec:cptheory3} details past CP methods for spatially referenced data. In Section \ref{sec:methods3}, our newly developed spatial CPs method is detailed and we develop a multivariate boundary detection spatial process for localized smoothing. We apply our method to a dataset of VF tests from glaucoma patients in Section \ref{sec:glaucoma3} and present a simulation study in Section \ref{sec:simstudy3}. We conclude in Section \ref{sec:disc3} with a discussion.


\section{Visual field data \label{sec:data}}

The visual ability of a glaucoma patient is most often measured through standard automated perimetry (SAP), an interactive technology that estimates the threshold sensitivity to light across a patient's field of vision. In this study, we analyze data acquired with the Humphrey Field Analyzer-II (HFA-II) (Carl Zeiss Meditec Inc., Dublin, CA), a machine that produces a visual acuity map across 54 testing locations on the VF. Two of these locations correspond to a natural blind spot corresponding to the optic disc, resulting in 52 informative points. During a test, the light is generated with increasing brightness ranging from approximately 40 decibels (dB) (excellent vision) to 0 dB (near blindness). The patient uses a hand-held trigger to indicate whether the light stimulus was detected and the machine reports the dimmest stimulus detected at each location. All stimuli that are not detected by 0 dB are censored due to the constraints of the machine, as values $<$ 0 dB are theoretically possible. The resulting gridded spatial layout represents an estimate of a patient's visual ability across the entire field of vision at that point in time. 

\begin{figure}[t] 
\begin{center}
\includegraphics[scale=0.75, trim = 0cm 0cm 0cm 3cm, clip]{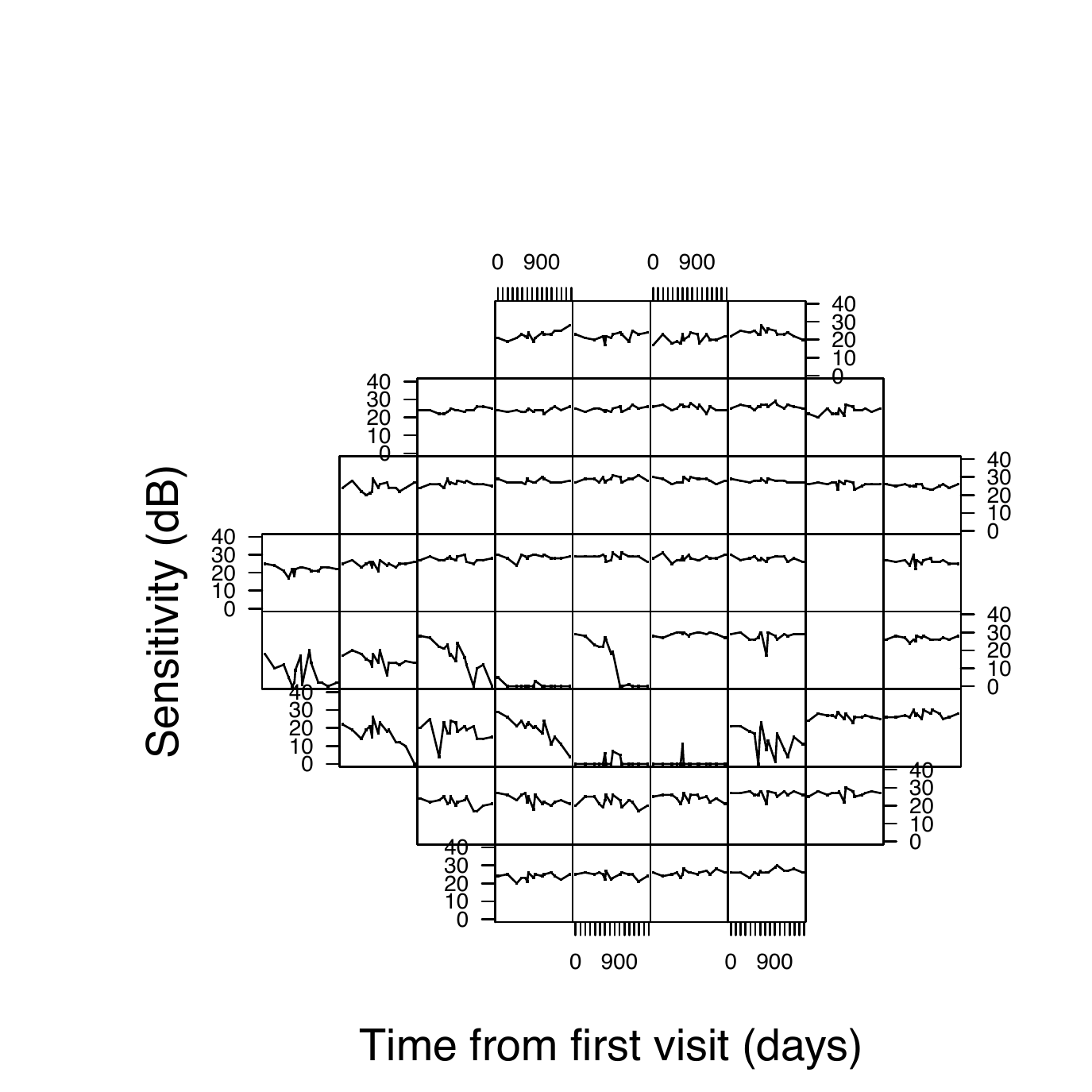}\\
\caption{Example visual field (VF) time series data for a patient with glaucoma. At each VF location, the time series of the observed threshold sensitivities is plotted. The trends and clustering of the time series motivate the use of spatially varying change points.  \label{fig:vfts}}
\end{center}
\end{figure}

In the course of monitoring a glaucoma patient, VF testing is performed on a regular basis and a longitudinal series of VFs is obtained. In Figure \ref{fig:vfts}, the time series of VF observations is presented for a selected patient from our data as a motivating example for the CP framework. The majority of past modeling attempts have assumed a linear relationship between VF sensitivities and time. However, from the data in Figure \ref{fig:vfts}, it is clear that CPs may provide an improved description of the trajectory of VF sensitivities over time, and in particular for areas of decreased sensitivities. Furthermore, the figure illuminates the inverse relationship between VF sensitivities and testing variability that is well known but rarely modeled in practice \citep{russell2012relationship}.

In order to fully motivate the modeling framework, we provide some details on the anatomy underlying the data generating mechanism for VF data. The light stimulus from the HFA-II is absorbed by RGCs, which make up the retinal nerve fiber layer (RNFL) of the retina, that use photoreceptors to transmit information to the brain along their axons \citep{davson2012physiology}. The RGCs disperse across the RNFL and converge at the optic disc, where the RGC axons bundle to form the optic nerve. For glaucoma patients, an increase in intraocular pressure creates stress on the optic disc, which in turn causes RGC death, interrupting the visual pathway to the brain. In particular, damage to specific regions of the optic disc corresponds to loss of RGCs whose axons enter the damaged region. 

This anatomy is important to understand when analyzing VF data, because vision is the brain's projection of information encoded by RGCs. Therefore, deterioration of the VF has a direct correspondence to damage of RGCs that originate from the same region of the retina. Because of this anatomy, correlation between two test points on the VF is dependent on the spatial proximity that their underlying RGCs enter the optic disc. This anatomical relationship was studied by \citet{garway2000mapping} who quantified the relationship between the VF and optic disc by estimating the angle that each test location's underlying RGC axons enters the optic disc, measured in degrees ($\circ$). The measure ranges from 0-360$^\circ$, where $0^\circ$ is designated at the 9-o'clock position (right eye) and angles are counted counter clockwise. Throughout this analysis, we utilize these angles to properly inform the spatial neighborhood and correlation structure of VF data, in which there is precedent \citep{betz2013spatial, zhu2014detecting, warren2016statistical, berchuck2018}. 

Our newly developed method is applied to data from the Vein Pulsation Study Trial in Glaucoma and the Lions Eye Institute trial registry, Perth, Western Australia. The dataset contains 1,448 VF tests from 194 distinct eyes (98 patients in total). All of the subjects have some form of primary open angle glaucoma. The mean follow-up time for participants is 934 days (2.5 years) with an average of 7.4 tests per subject. Every VF series is diagnosed as progressing based on the clinical judgment of two independent clinicians. In the case that the two clinicians disagree, a third clinician is consulted (occurred for only 13 VF series). In our study, we have 141 (74\%) stable and 50 (26\%) progressing patient eyes. These data have been previously analyzed \citep{betz2013spatial,warren2016statistical,berchuck2018}, in which additional details of the data can be found.


\section{Change point modeling for spatially referenced data \label{sec:cptheory3}}

VF data represent spatially referenced time series that are monitored for abnormalities that may indicate disease progression. Identifying abruptly changing variability in time series data was first characterized as a CP problem by \citet{Page1954} and was used as a quality control method to automatically detect faults in industrial processes. Although the CP literature is extensive \citep{quandt1958estimation,carlin1992hierarchical,julious2001inference}, the problem of detecting the timing of an unknown change in spatially referenced data is relatively sparse. The core of the literature involves extending the original CP model to account for spatial heterogeneity where the original CP model is given as
\begin{equation} \label{eq:cp}
\begin{split}
 Y_t=\left\{ \begin{array}{ll}
         {\beta}_0 + \beta_1 x_t + \epsilon_t& \quad \mbox{$t = 1,\ldots,\theta$},\\
         {\beta}_0 + \beta_1 x_{\theta} + {\beta}_2(x_t - x_{\theta}) + \epsilon_t & \quad \mbox{$t = \theta + 1,\ldots,\nu$}\end{array} \right.
\end{split}
\end{equation}
where $Y_t, \text{  } t=1,\ldots,\nu$, are observations of the time series, $x_t$ is a time varying regressor ($x_1 < \cdots < x_{\nu}$), and $\epsilon_t \stackrel{\text{iid}}{\sim}\text{N}\left(0, \sigma^2\right)$. The regression parameters $\beta_0$, $\beta_1$, and $\beta_2$ define a mean process that dictates a piece-wise linear regression separated at the CP, $\theta \in \{1,2,\ldots,\nu-1\}$, an index of the time series. 

With spatiotemporal data, the time series is now observed at a unique spatial location $\mathbf{s}$,  $\{Y_t(\mathbf{s}), t=1,\hdots,\nu\}$, across the interval $[x_1(\mathbf{s}), x_{\nu}(\mathbf{s})]$. Maximum likelihood theory has been developed for the standard CP method of Equation \ref{eq:cp} \citep{hinkley1970inference}, and its many extensions (e.g., multiple CPs), but more recently, and for spatial methods in particular, the Bayesian paradigm is preferred \citep{ulrich1981,carlin1992hierarchical}. There have been various Bayesian CP formulations that account for spatial or spatiotemporal dependencies through adjustments to the intercept \citep{beckage2007bayesian,minin2007phylogenetic,yu2008multilevel,smith2015change,cai2016bayesian} and extensions that allow for independent spatial processes before and after the CP \citep{majumdar2005spatio}. These methods introduce spatial dependency using conditional autoregressive (CAR) priors, a Gaussian Markov random field that induces neighborhood dependencies across the region \citep{besag1974spatial, geman1984stochastic}. 

In the context of Equation \ref{eq:cp}, this implies that spatial dependencies are most often modeled through adjustments to the intercepts, now spatially referenced as $\beta_0 + \beta_0(\mathbf{s})$. In particular, this implies the following conditional prior form, $\beta_0(\mathbf{s}_i) | \beta_0(\mathbf{s}_{j \neq i}), \tau^2 \stackrel{\text{ind}}{\sim} \text{N}\left(\sum_{j=1}^m w_{ij} \beta_0(\mathbf{s}_j) / w_{i+},  \tau^2 / w_{i+}\right)$, where $w_{ij} = 1(i \sim j)$ is an indicator equal to one if locations $i$ and $j$ are neighbors and $w_{i+} = \sum_{j = 1}^m w_{ij}$ is the number of neighbors for location $i$. This conditional prior distribution has an intuitive interpretation as the mean is an average of the neighboring values and the variance decreases as the number of neighbors increases \citep{banerjee2003hierarchical}. Although the CAR prior is an effective method for modeling spatial variability for areal data, these previous methods ignore the spatial variability in the CPs themselves.

In a recent study, \citet{wagner2014modeling} introduced a model with spatially varying intercepts, slopes, and CPs. In their method, there is a unique CP at each spatial location and additionally, the process is re-parameterized so that the CP parameter is continuous. The general form of the mean structure (after applying the appropriate link function) is given as
\begin{align} \label{eq:sp2}
\mu_t(\mathbf{s})&=\left\{ \begin{array}{ll}
        {\beta}_0(\mathbf{s}) + \beta_1(\mathbf{s}) x_t(\mathbf{s}) & \text{ } \mbox{$x_t(\mathbf{s}) \leq \theta(\mathbf{s})$},\\
        {\beta}_0(\mathbf{s}) + \beta_1(\mathbf{s}) \theta(\mathbf{s}) + {\beta}_2(\mathbf{s})\{x_t(\mathbf{s}) - \theta(\mathbf{s})\}& \text{ } \mbox{$x_t(\mathbf{s}) > \theta(\mathbf{s}).$}\end{array} \right.
\end{align}
In this specification, the CP $\theta(\mathbf{s})$ is no longer an index of the time series and instead ranges in $[x_1(\mathbf{s}), x_{\nu}(\mathbf{s})]$. 
In \citet{wagner2014modeling}, after rescaling the CPs to the real line, the regression parameters and CPs are modeled jointly using a multivariate Gaussian distribution centered at a vector of fixed effects with an unstructured cross-covariance matrix. The authors assumed independence across space in their model specification. Subsequently, this method has been extended to account for spatial dependency using a multivariate CAR (MCAR) process \citep{warren2017spatial}. The MCAR allowed for an improved modeling framework by accounting for both spatial variability and cross-covariance in the location specific intercepts, slopes, and CPs.


\section{Methods \label{sec:methods3}}

Following the approach of \citet{warren2017spatial}, we propose extending the general CP framework by introducing a set of spatially varying intercepts, slopes, and CPs that are modeled with an MCAR process. This increased flexibility allows us to accurately describe the individual time series behavior at each unique spatial location while accounting for spatial correlation and cross-covariance between the parameters that may result in increased power to detect important changes.   

However, we extend the existing framework in a number of ways.  First, we also allow for a CP (and spatially varying regression coefficients) in the model for the nuisance (or variance) parameter to reflect the situation where the observed data become more or less variable after a CP occurs. Next, we define the observed CP at a specific spatial location as a function of a potentially censored latent CP process (also spatially varying). This flexible specification permits spatial interpolation of currently censored or unobserved CPs across the spatial domain. This becomes particularly useful when working with data where the CPs describe the timing of treatment failure or an adverse event. Finally, a novel boundary detection MCAR process is developed that allows for a dissimilarity metric in the spatial covariance definition. Inference for this model is based on Markov chain Monte Carlo (MCMC) simulation, and a description of the algorithm is given in \ref{sec:mcmcappndx}. The newly introduced model is computationally intensive, so the MCMC algorithm is implemented using Rcpp \citep{eddelbuettel2011rcpp} and is available from the R package \texttt{spCP} \citep{Rcore}. 


\subsection{Spatially varying change points model \label{sec:spcpmodel}}

We begin by describing our new methodology generally before applying it to VF data in Section \ref{sec:glaucoma3}. The general framework is flexible enough to accommodate many different spatially referenced time series data settings. Let $Y_t\left(\mathbf{s}\right)$ be an observation from spatial location $\mathbf{s}$ at time $x_t\left(\mathbf{s}\right)$, for $\{\mathbf{s}_i:i = 1,\ldots m\},\ t = 1,\ldots,\nu$, where the times are strictly increasing. An extension of Equation \ref{eq:sp2}, we introduce a model that allows for location specific CPs, intercepts, and slopes in the mean and nuisance parameter processes of the data such that
\begin{align} \label{eq:spmean} \notag
&Y_t(\mathbf{s}) | \mu_t(\mathbf{s}), \boldsymbol{\zeta}_t(\mathbf{s}) \stackrel{\text{ind}}{\sim}{} f\left(\mu_t(\mathbf{s}), \boldsymbol{\zeta}_t(\mathbf{s})\right), \text{ } g_1\{\mu_t(\mathbf{s})\} = \gamma_t(\mathbf{s}), \text{ } g_2\{\zeta_t(\mathbf{s})\} = \omega_t(\mathbf{s})\\
&\gamma_t(\mathbf{s})=\left\{ \begin{array}{ll} {\beta}_0(\mathbf{s}) + {\beta}_1(\mathbf{s}) x_t(\mathbf{s}) & x_t(\mathbf{s}) \leq \theta_1(\mathbf{s})\\ {\beta}_0(\mathbf{s}) + {\beta}_1(\mathbf{s}) \theta_1(\mathbf{s}) + {\beta}_2(\mathbf{s})\left\{x_t(\mathbf{s})-\theta_1(\mathbf{s})\right\} & x_t(\mathbf{s}) > \theta_1(\mathbf{s})\end{array} \right.\\
&\omega_t(\mathbf{s})=\left\{ \begin{array}{ll} {\lambda}_0(\mathbf{s}) + {\lambda}_1(\mathbf{s}) x_t(\mathbf{s}) & x_t(\mathbf{s}) \leq \theta_2(\mathbf{s})\\ {\lambda}_0(\mathbf{s}) + {\lambda}_1(\mathbf{s}) \theta_2(\mathbf{s}) + {\lambda}_2(\mathbf{s})\left\{x_t(\mathbf{s})-\theta_2(\mathbf{s})\right\} & x_t(\mathbf{s}) > \theta_2(\mathbf{s}).\end{array} \right. \notag
\end{align}
The form of Equation \ref{eq:spmean} is an extension of Equation \ref{eq:sp2} with a general likelihood, $f$, and a CP in the nuisance process. The mean process is modeled through a link function, $g_1(\cdot)$, and thus can be non-linear. The nuisance parameters are partitioned into constant terms and a location and time specific component, such that $\boldsymbol{\zeta}_t(\mathbf{s}) = \{\zeta_t(\mathbf{s}), \boldsymbol{\zeta}\}$. The CP for the nuisance parameter is introduced similarly as for the mean process, through the link function $g_2(\cdot)$. At each spatial location, the CP for the mean process, $\theta_1(\mathbf{s})$, and nuisance parameter process, $\theta_2(\mathbf{s})$, are unique.

Allowing for a continuous valued CP is important because we generally do not expect the change to occur discretely during one of the observed time points, particularly when there is a large amount of time between observations. The specification can be improved however, through a natural assumption about the censoring of CPs at the edges of the observed time period. In particular, when $\theta_j(\mathbf{s}) = x_1(\mathbf{s})$, $j=1,2$, the CP has occurred before the time series at location $\mathbf{s}$ was monitored, while $\theta_j(\mathbf{s}) = x_{\nu}(\mathbf{s})$ indicates that the CP has not yet occurred by the end of the follow-up period. As such, both the lower and upper bounds of the CP are censored. To account for this, we model the observed CP as a function of the underlying and potentially unobserved true latent CP process, $\eta_j\left(\mathbf{s}\right)$, such that $\theta_j(\mathbf{s}) = \max\{\min\{\eta_j(\mathbf{s}),x_{\nu}(\mathbf{s})\},x_1(\mathbf{s})\}$ where $\eta_j(\mathbf{s})$ has support on the real line and is allowed to vary spatially. This framework gives us the ability to spatially interpolate latent CP parameters that have yet to occur which could be important in some applications for determining when and where future damage can be expected. 
 

\subsection{Multivariate dissimilarity metric spatial process \label{sec:mcar}}

In order to account for the potentially complex spatial correlation and cross-covariance between the introduced spatially varying intercepts, slopes, and CPs, we develop a novel boundary detection MCAR model that allows for a dissimilarity metric within the framework introduced by \citet{mardia1988multi}. Incorporating the dissimilarity metric into the framework is important for correctly modeling areal spatial data where the neighborhood structure of the region is uncertain and controlled by a set of underlying factors (e.g., VF data).  Similar to the CAR process, the MCAR introduces spatial dependencies through a Markov random field technique, and is in fact a multivariate generalization of the CAR with a seperable covariance. Since \citet{mardia1988multi}, the MCAR has seen numerous developments throughout the years that have generalized the seperable assumption \citep{gelfand2003proper,jin2005generalized,jin2007order}. Although there may be motivation for a non-seperable process, in this paper we introduce our boundary detection technique within the standard MCAR. This could easily be generalized to a non-seperable process using the linear model of coregionalization. 

Similar to the univariate case, the joint distribution of the parameters is derived from their full conditional distributions. Under the Markov random field assumption, these conditional distributions can be specified as follows,
\begin{equation} \notag
\begin{split}
f\left(\boldsymbol{\phi}_i|\boldsymbol{\phi}_{j\neq i}, \boldsymbol{\mu},\boldsymbol{\Gamma}_i\right) = \text{MVN}\left(\boldsymbol{\mu}_i + \sum_{i \sim j} \mathbf{B}_{ij} (\boldsymbol{\phi}_j - \boldsymbol{\mu}_j) , \boldsymbol{\Gamma}_i\right), \quad i,j = 1, \ldots, m
\end{split}
\end{equation}
where $\boldsymbol{\phi}_i = (\phi_{i1}, \phi_{i2}, \ldots, \phi_{ip})^{\text{T}}$ and $\boldsymbol{\mu}_i = (\mu_{i1},\mu_{i2},\ldots,\mu_{ip})^{\text{T}}$ are $p$-dimensional vectors, $\boldsymbol{\mu} = (\boldsymbol{\mu}_1^{\text{T}}, \boldsymbol{\mu}_2^{\text{T}},\ldots,\boldsymbol{\mu}_{m}^{\text{T}})^{\text{T}}$, and $\boldsymbol{\Gamma}_i$ and $\mathbf{B}_{ij}$ are $p \times p$ matrices. Using Brook's lemma \citep{brook1964distinction}, Mardia proved that the full conditional distributions uniquely determine the joint distribution given by $f\left(\boldsymbol{\phi}|\boldsymbol{\mu},\{\boldsymbol{\Gamma}_i\}\right) \propto \exp\left\{-\frac{1}{2}(\boldsymbol{\phi} - \boldsymbol{\mu})^{\text{T}}\boldsymbol{\Gamma}^{-1}(\mathbf{I}_{mp} - \tilde{\mathbf{B}})(\boldsymbol{\phi} - \boldsymbol{\mu})\right\},$
where $\boldsymbol{\phi} = (\boldsymbol{\phi}_1^{\text{T}}, \boldsymbol{\phi}_2^{\text{T}}, \ldots, \boldsymbol{\phi}_m^{\text{T}})^{\text{T}}$, $\boldsymbol{\Gamma}$ is a block diagonal matrix with blocks $\boldsymbol{\Gamma}_i$ and $\tilde{\mathbf{B}}$ is an $mp \times mp$ matrix with $(i,j)$-th block $\mathbf{B}_{ij}$. Different MCAR forms can be obtained through various $\boldsymbol{\Gamma}$ and $\tilde{\mathbf{B}}$.

In this case, we induce an MCAR structure that generalizes the univariate version of the Leroux CAR specification, thus allowing the dissimilarity metric to be used in a multivariate setting \citep{leroux2000estimation,macnab2007regression,lee2011boundary}. As in the univariate case, symmetry of $\boldsymbol{\Gamma}^{-1}(\mathbf{I}_{mp} - \tilde{\mathbf{B}})$ is required. A convenient special case sets $\mathbf{B}_{ij} = b_{ij}\mathbf{I}_p$. Under this condition, the multivariate version of the Leroux dissimilarity metric likelihood can be obtained with the following specifications, $b_{ij} = \rho w_{ij}\left(\alpha\right) \Delta^{-1}$ and $\boldsymbol{\Gamma}_i = \boldsymbol{\Sigma} \Delta^{-1}$, where $\Delta = \rho\sum_{j=1}^m w_{ij}\left(\alpha\right) + (1-\rho)$. Spatial neighborhood structure is introduced through the adjacencies $\{w_{ij}(\alpha)\}$. 

The adjacencies are defined similar to the specification in \citet{berchuck2018}. In particular, they are a function of a dissimilarity metric and regression parameter such that $w_{ij}(\alpha) = 1\{i\sim j\}\exp\{-z_{ij}\alpha\}$ where $i\sim j$ is the event that locations $i$ and $j$ share an edge or corner ($w_{ii} = 0$ for all $i$). The quantity $z_{ij}$ is a measure of (dis-)similarity between two locations and is defined as $z_{ij} = |z_i - z_j|$, where $z_i$ is a covariate at location $\mathbf{s}_i$. The parameter $\alpha$ is forced to be non-negative, so that the adjacencies are in the open unit interval, and this is in fact a condition for guaranteeing a valid covariance. 

This specification induces a multivariate version of the conditional distributions which is analogous to the univariate Leroux likelihood and is given as $f\left(\boldsymbol{\phi}_i|\boldsymbol{\phi}_{j\neq i}, \boldsymbol{\delta},\boldsymbol{\Sigma}, \alpha,\rho\right) =$
\begin{equation} \notag
\begin{split}
 \text{MVN}\left(\frac{\rho\sum_{j=1}^m w_{ij}\left(\alpha\right) \boldsymbol{\phi}_j + (1 - \rho) \boldsymbol{\delta}}{\rho \sum_{j=1}^m w_{ij}\left(\alpha\right) + (1-\rho) } , \frac{\boldsymbol{\Sigma}}{\rho\sum_{j=1}^m w_{ij}\left(\alpha\right) + (1-\rho)}\right)
\end{split}
\end{equation}
where $\boldsymbol{\mu}_i = \boldsymbol{\delta}$, so that the mean is constant across spatial locations. The distribution for the joint specification can be established by noting that $\boldsymbol{\Gamma}^{-1}(\mathbf{I}_{mp} - \tilde{\mathbf{B}}) = \mathbf{Q}\left(\alpha, \rho\right)
 \otimes \boldsymbol{\Sigma}^{-1},$ with $\mathbf{Q}(\alpha, \rho) = \rho\mathbf{W}^*(\alpha) + (1-\rho)\mathbf{I}_m$. The matrix $\mathbf{W}^*\left(\alpha\right)$ has diagonal elements $w^*_{ii}\left(\alpha\right)=\sum_{j=1}^m w_{ij}\left(\alpha\right)$ and off-diagonal elements $w^*_{ij}\left(\alpha\right)=-w_{ij}\left(\alpha\right)$. 
Finally, this form of the adjacencies yields the joint distribution $f\left(\boldsymbol{\phi}|\boldsymbol{\delta},\boldsymbol{\Sigma}, \alpha, \rho\right) \sim \text{MVN} \left(\mathbf{1}_m \otimes \boldsymbol{\delta}, \mathbf{Q}\left(\alpha, \rho\right)^{-1}
 \otimes \boldsymbol{\Sigma}\right)$. We denote a random variable $\boldsymbol{\phi}$ with this distribution as $\boldsymbol{\phi} \sim \text{MCAR}(\alpha, \rho, \boldsymbol{\delta},\boldsymbol{\Sigma})$. 
  

\subsection{Hyperprior distributions \label{sec:hypers3}}

To complete the model specification, we select prior distributions for the hyperparameters. The prior for the mean $\boldsymbol{\delta}$, covariance $\boldsymbol{\Sigma}$, and boundary detection parameter $\alpha$ are as follows;
\begin{equation} \notag
\begin{split}
\begin{array}{ccc}
\boldsymbol{\delta} \sim \text{N}\left(\mathbf{0},\kappa^2\mathbf{I}_p\right), & \boldsymbol{\Sigma} \sim \text{Inverse-Wishart}(\xi, \boldsymbol{\Psi}), & \text{and }  \alpha \sim \text{Uniform}\left(a_{\alpha}, b_{\alpha}\right).\\
\end{array}
\end{split}
\end{equation}
The mean $\boldsymbol{\delta}$ is zero centered with variance, $\kappa^2$, which is chosen to be large in order to yield a weakly informative prior. For the prior on $\boldsymbol{\Sigma}$, we specify an inverse-Wishart distribution with degrees of freedom $\xi=p+1$ and scale matrix, $\boldsymbol{\Psi}=\mathbf{I}_{p}$, where $p$ is the dimension of the spatial parameter, $\boldsymbol{\phi}_i$. This prior induces marginally uniform priors on the correlations of $\boldsymbol{\Sigma}$ and allows for the diagonals to be weakly informative \citep{gelman2014bayesian}. The dissimilarity metric parameter, $\alpha$, is forced to be non-negative with a uniform distribution. The bounds are defined as $a_{\alpha} = 0$ and $b_{\alpha} = -\log\{0.5\}/\min_{i,j}\{z_{ij}\}$. As in \citet{lee2011boundary}, the upper limit is set so that the smallest dissimilarity metric between two locations can still obtain a spatial adjacency of 0.50.

We fix $\rho$ at $0.99$ to force spatial dependency, in which there is precedent \citep{lee2011boundary,berchuck2018}. In particular, \citet{berchuck2018} demonstrated that $\rho$ fixed at 0.99 has little impact on estimation of $\alpha$. Multivariate spatial structure is introduced through $\boldsymbol{\phi}_i = \left\{\beta_0\left(\mathbf{s}_i\right), \beta_1\left(\mathbf{s}_i\right), \beta_2\left(\mathbf{s}_i\right),\eta_1\left(\mathbf{s}_i\right), \lambda_0\left(\mathbf{s}_i\right), \lambda_1\left(\mathbf{s}_i\right), \lambda_2\left(\mathbf{s}_i\right),\eta_2\left(\mathbf{s}_i\right)\right\}^{\text{T}}$, which are assigned an MCAR($\alpha, 0.99, \boldsymbol{\delta}, \boldsymbol{\Sigma}$) process. 


\section{Monitoring visual field progression\label{sec:glaucoma3}}


\subsection{Change points model for visual field data \label{sec:cpvf}}

We modify the general form of the model described in Section \ref{sec:methods3} for the purpose of monitoring a patient's VF data across space and time with the goal of improving disease management. Using the notation developed in Section \ref{sec:methods3}, we define $Y_t^*\left(\mathbf{s}\right)$ as the observed VF sensitivity at VF location $\mathbf{s}$ and $x_t$ days after the baseline visit ($x_1 = 0$), for $\{\mathbf{s}_i:i = 1,\ldots 52\}, t = 1,\ldots,\nu$, where the number of visits, $\nu$, is patient specific. Note that since the timing of visits is the same for all VF locations for a single patient, $x_t(\mathbf{s}) \equiv x_t$. Furthermore, the number of VF locations, 52, represents the total number of locations on the VF after removing the two corresponding to the blind spot. 

The data generating mechanism for VF sensitivities is such that observations are censored at zero due to limitations of the machine that collects the data. Therefore, we use a Tobit model \citep{tobit1958estimation}, in which there is precedent in the glaucoma progression literature \citep{betz2013spatial,bryan2015global,berchuck2018}. To induce the Tobit model, we define a latent process through the standard Tobit link, $Y_t^*(\mathbf{s}) = \max\left\{0,Y_t(\mathbf{s})\right\}$, where $Y_t(\mathbf{s}) = \mu_t(\mathbf{s}) + \epsilon_t(\mathbf{s}), \epsilon_t(\mathbf{s})\stackrel{\text{ind}}{\sim}\text{N}\left(0, \sigma_t^2(\mathbf{s})\right)$. This is a special case of Equation \ref{eq:spmean} where $\zeta_t(\mathbf{s}) = \sigma_t^2(\mathbf{s})$, $g_1(\cdot)$ is the identity link, and $g_2(\cdot) = \log\{\sqrt{\{\cdot\}}\}$.

To appropriately model the course of glaucoma progression, we modify the mean and variance structures. Patients with glaucoma are typically stable until the point of progression when their VF time series begin to deteriorate (see Figure 1 for example). As such, for VF data we define the mean process so that at times before the CP there is a constant intercept at each VF location.  To accomplish this, we set $\beta_1(\mathbf{s}) = 0$ in Equation 3. The same assumption is used for the variance process, so that the variance can increase with the onset of deterioration (i.e., $\lambda_1(\mathbf{s}) = 0$). Furthermore, motivated by the disease course, we force the CPs in the mean and variance to occur simultaneously, $\theta(\mathbf{s})\equiv\theta_1(\mathbf{s}) = \theta_2(\mathbf{s})$.  To finalize the model, we specify hyperpriors based on details in Section \ref{sec:hypers3}, where $\kappa^2 = 1000$ and $p=5$. We define the dissimilarity metric, $z_i$, as the Garway-Heath angle at VF location $\mathbf{s}_i$.

We draw samples from the joint posterior by utilizing MCMC techniques \citep{metropolis1953equation,geman1984stochastic,gelfand1990sampling}. Many of the parameters have conjugate full conditionals. In particular, all of the parameters have closed form solutions, with the exception of $\alpha$, $\lambda_0(\mathbf{s})$, $\lambda_2(\mathbf{s})$, and $\eta(\mathbf{s})$. A Gibbs sampler is used with Metropolis steps for parameters lacking conjugacy. For each eye, the MCMC sampler is run for 250,000 iterations after a 10,000 burn-in and thinned to a final sample size of 10,000. There were no obvious signs of non-convergence observed for any of the 191 VF series based on traceplots and the Geweke statistic \citep{geweke1992}. The MCMC sampler is implemented in the R package \texttt{spCP}. More details about the MCMC sampler can be found in \ref{sec:mcmcappndx}.


\subsection{Model comparison \label{sec:mc}}

In order to establish the capability of the introduced model to predict VF sensitivities, predict the location and timing of future vision loss, and diagnose glaucoma progression, we compare it to a number of competing models with varying levels of complexity:
\begin{enumerate}
\item[1.] \emph{Point-wise linear regression (PLR)}: Independent Tobit linear regression at each location.
\item[2.] \emph{Non-spatial CP (D)}: Tobit linear regression with \textbf{discrete} CPs, based on Equation \ref{eq:cp}.
\item[3.] \emph{Non-spatial CP (C)}: Tobit linear regression with \textbf{continuous} CPs, where the CPs at each VF location are independent and uniformly distributed a priori.
\item[4.] \emph{Non-spatial CP (L)}: Tobit linear regression with continuous CPs and an underlying normally distributed \textbf{latent} process, as specified in Section \ref{sec:spcpmodel}.
\item[5.] \emph{Spatial CP}: Spatial CPs model detailed in Section \ref{sec:cpvf}.
\end{enumerate}
Each of the competing CP models specify a constant process before the CP, as specified in Section \ref{sec:cpvf}. In addition, each of the models utilize the Tobit likelihood in order to gain an understanding of key differences in their performance due to other factors. The motivation of fitting the simplified models, such as PLR and basic CP models, is to have a benchmark for comparison and to emphasize the importance of the latent CP specification and spatial structure. Note that each of the models are subsets of Spatial CP.

The models are compared using model fit diagnostics and hold-out prediction criterion. To assess model fit, the deviance information criterion (DIC) is used \citep{spiegelhalter2002bayesian}. For prediction, the final visit for each of the VF series is held out, and the remaining visits are used to predict the hold-out visit. Prediction accuracy is measured using the mean squared prediction error (MSPE). Smaller values of DIC and MSPE are preferred.

The model fit and prediction diagnostics can be found in Table \ref{tab:diagpred} averaged over all patients. The DIC for Spatial CP is superior to the rest of the models, with a minimal value of -53.74. Generally, there is a monotone decreasing trend in DIC as the models become more complex. The number of effective parameters also trends smaller with added model complexity, indicating that the added model structure provides a more concise framework for modeling VF data. The prediction measure also favors Spatial CP, with optimal MSPE of 0.13. MSPE for the three non-spatial CP models is inferior to PLR, indicating that the inclusion of a CP does not guarantee improved prediction. Only when accounting for the spatial nature of VF data does a CP model offer improved prediction over PLR, again illuminating the importance of modeling spatial variability. Overall, the results in Table \ref{tab:diagpred} indicate that Spatial CP has superior model fit and prediction ability for VF data.

\setlength{\tabcolsep}{4pt}
\begin{table}[t]
\centering
\caption{Model fit and prediction diagnostics for model comparison. Model fit is determined using the deviance information criterion (DIC) and effective number of parameters ($p_D$). Prediction capability is determined using the mean squared prediction error (MSPE). The ability of each model to diagnose progression is assessed using a metric dependent on the latent change points (CPs). For each model with a latent CP process, the metric $\max_{i}\{P[\theta(\mathbf{s}_i) < t_{\nu}|\mathbf{Y}]\}$ is calculated and regressed against the clinically determined progression status. Diagnostic capability is measured using Akaike information criterion (AIC), area under the receiver operating characteristic curve (AUC), and the p-value for each measure.\label{tab:diagpred}} 
\begin{tabular}{lrrcccrrr}
  \hline
& \multicolumn{2}{c}{Model Fit} & & \multicolumn{1}{c}{Prediction} & & \multicolumn{3}{c}{Diagnostic} \\ \cline{2-9}
Model & DIC & $p_D$ & & MSPE & & AIC & AUC & p-value\\  \hline
PLR & 31.11 & 120.30 & & 0.20 & & \multicolumn{1}{c}{---}& \multicolumn{1}{c}{---}&\multicolumn{1}{c}{---} \\ 
  Non-spatial CP (D) & -14.27 & 119.30 & & 0.45 & &\multicolumn{1}{c}{---} & \multicolumn{1}{c}{---}& \multicolumn{1}{c}{---}\\ 
  Non-spatial CP (C) & -18.76 & 112.13 & & 0.32 & & \multicolumn{1}{c}{---}& \multicolumn{1}{c}{---}& \multicolumn{1}{c}{---}\\ 
  Non-spatial CP (L) & -29.05 & 67.69 & & 0.23 & & 211.36 & 0.64 & 0.001  \\ 
  Spatial CP & -53.74 & 68.21 & & 0.13 & & 208.67 & 0.69 & $<$0.001\\ \hline
  \end{tabular}
\end{table}


\subsection{Diagnosing progression using change points \label{sec:diagcps}}

To illustrate the superiority of the introduced methodology, in addition to model fit and prediction, we demonstrate the diagnostic capability of the CPs. We hypothesize that the location of the CPs can serve as a proxy for disease progression and propose a diagnostic metric that is a function of the posterior probability that a CP has occurred up to the last VF visit, $p_i = P[\theta(\mathbf{s}_i) < t_{\nu}|\mathbf{Y}]$. We define the final metric for a patient eye as the maximum across all VF probabilities, $\max\{p_i: i=1,\hdots, 52\}$. This is intuitive because a longitudinal series of VFs with earlier occurring CPs is more likely to be progressing. 

In order to assess the diagnostic ability of the CPs, we construct logistic regression models, regressing the defined metric on the clinical assessment of progression and present Akaike information criterion (AIC), area under the receiver operating characteristic curve (AUC), and a p-value for the slope parameter. These measures are presented in Table \ref{tab:diagpred} for models that have a latent CP process (the only models capable of estimating the CP probability). From Table \ref{tab:diagpred}, it is clear that the CPs arising from Spatial CP are highly associated with progression (p-value: $<0.001$). The non-spatial model also yields a significant slope estimate, however the p-value is of a different magnitude (p-value: $0.001$). Furthermore, the values of AIC and AUC are optimized under Spatial CP with minimal AIC, 208.67, and maximal AUC, 0.69. These results help validate the claim that the CPs are representative of disease progression and that their estimates are clinically useful.

\begin{figure} 
\begin{center}
\includegraphics[scale=0.60, trim = 1.5cm 0cm 0cm 0cm, clip]{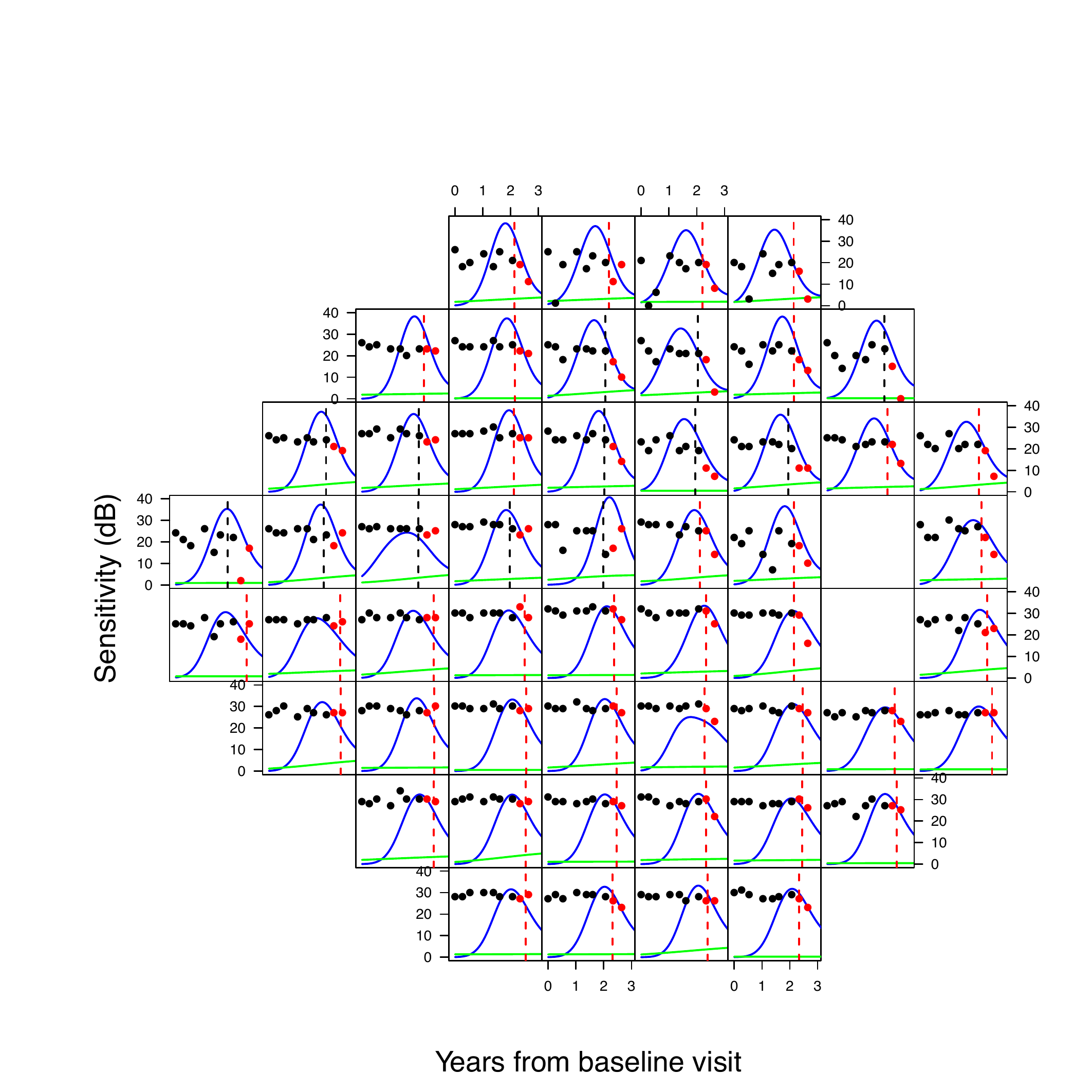}\\
\caption{Demonstrating the ability of Spatial CP to identify change points (CPs) that have yet to occur. At each cell, the observed DLS are presented for an example patient with nine visits. Based on only the first seven visits, the posterior density estimates of the CPs for Spatial CP and Non-spatial CP (L) are presented as density estimates in blue and green lines, respectively. The estimated CP based on Spatial CP using all of the nine VF visits are presented as dashed vertical lines. The DLS and CPs are shown in red if they occur after the seventh visit and thus are not observable at the time the posterior densities were estimated.  \label{fig:extra}}
\end{center}
\end{figure}

To further illustrate the diagnostic ability granted by the latent CP framework, we present an example patient as a proof of concept. This patient's information is appropriate for investigation because they have late occurring CPs that are truncated when removing visits near the end of follow-up. The patient has nine VF visits, however we only use the initial seven to fit Spatial CP and spatially interpolate the latent CPs. We then compare these predictions with the actual CPs estimated using the full set of data (i.e., all nine visits). The patient's VF data are presented in Figure \ref{fig:extra}, where at each location the first seven VF sensitivities are shown using black dots and the last two are shown in red. Furthermore, the mean CPs estimated by Spatial CP for the full data are presented as dashed vertical lines. Any CP that was estimated to have occurred after the seventh visit is colored red and is technically truncated for purposes of clinical use at visit seven. Finally, the posterior latent CP density estimates (based on the reduced data) for Spatial CP and Non-spatial CP (L) are presented as density estimates in blue and green, respectively.

From Figure \ref{fig:extra} it is clear that Spatial CP is capable of predicting future CPs that have not been observed. This is remarkable, since at the seventh visit there was no clear sign of a CP at many of these locations and furthermore, Non-spatial CP (L) performed poorly, indicating the importance of spatial structure. To quantify the conclusions drawn from Figure \ref{fig:extra}, for each model we present diagnostics for prediction of the estimated CPs. Each model is compared to the ``true" CPs estimated from the same model with full data. Spatial CP and Non-spatial (L) have MSPE of 4.70 and 8,525.82, respectively, indicating Spatial CP provides superior prediction of the unobserved CPs. These powerful results indicate that the introduced methodology has the capability of predicting future unobserved CPs, and therefore progression. This is possible due to both the latent CP specification and multivariate spatial boundary detection structure. As shown in Table \ref{tab:diagpred} and Figure \ref{fig:extra}, without either of these specifications the model suffers in terms of model fit, prediction, and diagnostic capability. 


\subsection{Clinical implementation}
\label{sec:cu}

\begin{figure} 
\begin{center}
\includegraphics[scale=1.1]{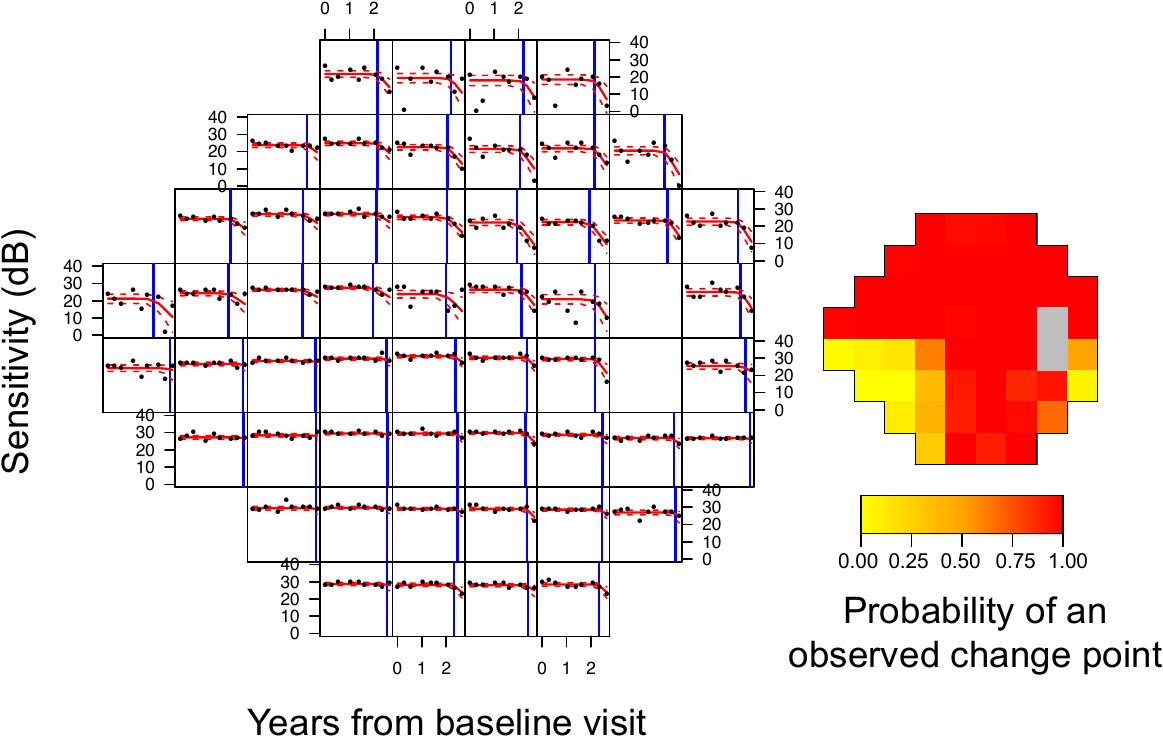}\\
\caption{Output generated from Spatial CP. The plot on the left presents the estimated mean (with 95\% credible intervals) and CPs for the same example patient in Figure 1. The mean process is presented in red and the change points are vertical blue lines. The panel on the right presents the posterior probability of an observed CP at each location on the VF. \label{fig:vfoutput}}
\end{center}
\end{figure}

Having established the potential for using Spatial CP in the clinical setting for prediction of VF sensitivities and CPs, we now present example output for use clinically. In Figure \ref{fig:vfoutput}, on the left, the posterior fitted trend is presented on the VF with the estimated posterior CPs. The plot presents the estimated regression mean (with 95\% credible intervals) and CPs for the same example patient in Figure \ref{fig:vfts}. The mean process is presented in red and the CPs are vertical blue lines. This figure demonstrates the flexibility of Spatial CP to produce non-linear trends, not simply two lines connected at a point. This type of trajectory is a result of the CPs being unknown quantities. In addition to the predicted fit, it is useful to have a heatmap of the posterior probability that each of the CPs have occurred in the observed data (right panel of Figure \ref{fig:vfoutput}). This presentation corresponds to the estimated CPs in the left frame, but further provides a useful map for clinicians to understand the location and severity of progression. 

Both of these plots in conjunction with the diagnostic potential from Spatial CP provide a useful framework for clinicians to assess progression. This presentation is clinically useful, and can be further improved when presented in video format. In Video \ref{vid:gif}, the heatmap of Figure \ref{fig:vfoutput} is presented over the full course of follow-up and into the future to provide clinicians an understanding of the disease pattern and growth over time and into the future. The combination of the latent CP process and novel spatial structure has produced a technique that goes beyond the simple CP models and PLR in terms of model fit, prediction, and progression diagnoses. 


\section{Simulation study \label{sec:simstudy3}}

The purpose of the simulation study is to investigate the performance of Spatial CP and the comparison models detailed in Section \ref{sec:mc} under different data generating scenarios. The scenarios are all subsets of the joint spatial structure, $\text{MVN} \left(\mathbf{1}_m \otimes \boldsymbol{\delta}, \mathbf{Q}\left(\alpha, 0.99\right)^{-1} \otimes \boldsymbol{\Sigma}\right)$, with parameters chosen to reflect those from an actual analyzed glaucoma patient:
\begin{equation} \label{eq:simsettings}
\begin{split}
\boldsymbol{\delta} = \left[\begin{array}{r}
 25 \\ 
-15 \\ 
-0.5 \\ 
 0.1 \\ 
 0.5  \end{array} \right], \text{ } \boldsymbol{\Sigma} = \left[\begin{array}{rrrrr}
 0.025 & -0.500 & -0.500 & -0.500 &  0.500 \\ 
-0.500 &  0.025 &  0.500 &  0.500 & -0.500 \\ 
-0.500 &  0.500 &  0.025 &  0.250 & -0.500 \\ 
-0.500 &  0.500 &  0.250 &  0.025 & -0.500 \\ 
 0.500 & -0.500 & -0.500 & -0.500 &  0.025 \end{array} \right],
\end{split}
\end{equation}
and $\alpha = 0.10$. Based on these true parameters, we create five data generating settings: (1) Progressing: CPs have all occurred before the beginning of follow-up; (2) Stable: CPs have yet to occur; (3) Non-spatial CPs: CPs are observed throughout follow-up in the mean and variance process with no spatial dependency; (4) Spatial CPs with no covariance: spatially dependent CPs are observed throughout follow-up in the mean and variance process with no cross-covariance in the spatial effects; (5) Full spatial CP model: spatially dependent CPs are observed throughout follow-up in the mean and variance process with spatial cross-covariance. 

Settings 1-4 are obtained by changing components of $\boldsymbol{\delta}$, $\boldsymbol{\Sigma}$, and $\alpha$. For Settings 1 and 2, $\alpha$ is fixed at an arbitrarily large value, $1000$, to remove spatial dependency, and $\boldsymbol{\Sigma}$ is restricted to being diagonal, with the fourth entry equal to zero to remove a CP in the variance. This specification yields the PLR model as long as the CPs are not observed. As such, Setting 1 is obtained by fixing the fifth entry in $\boldsymbol{\delta}$ to an arbitrarily large negative value, -10, while a large positive value, 10, is used for Setting 2. To guarantee no CP in the variance, the fourth entry in $\boldsymbol{\delta}$ is fixed at zero. Setting 3 introduces a CP in the variance process by using the specification of $\boldsymbol{\delta}$ from Equation \ref{eq:simsettings} and the fully diagonal $\boldsymbol{\Sigma}$. The value of $\alpha$ is still fixed at 1000, so that there is no spatial dependence. Setting 4 is obtained by specifying $\boldsymbol{\Sigma}$ to be diagonal and represents a spatial CP model with no cross-covariance between spatially referenced effects. Setting 5 uses the full parameter specification.

A new dataset is generated for each simulated vector of $\boldsymbol{\phi}$ to ensure that the results are not affected by a particular realization. For each setting, we simulate data based on a set of known hyperparameters and then fit each model to the same dataset to assess model fit, prediction, and estimation of CPs (both latent and observed). As in Section \ref{sec:glaucoma3}, for model fit we utilize DIC and $p_D$ for exploring explanatory ability of each model, and MSPE for prediction. The framework for assessing prediction is innately linked to how the data were simulated. Each dataset was simulated assuming there were 21 VF visits (the maximum number of visits in our study data), however, the models were only fit to the first 14 visits. This allows us to use the last 7 visits for prediction assessment. The visit times were specified as the sequence of numbers between 0 and 1 by a 0.05 increment. Finally, estimation of both the observed CPs and latent CPs process are assessed using bias, MSE, and empirical coverage (EC) of the Bayesian credible intervals. In this context, the EC is the proportion of the simulated datasets where the calculated 95\% credible interval contains the true parameter value.

The results from the study are displayed in Tables \ref{tab:sim1}, \ref{tab:sim2}, and \ref{tab:sim3} and relate to 250 simulated data sets for each setting. In Table \ref{tab:sim1}, the model fit diagnostics, DIC, and $p_D$, are presented for the five models across all data generating settings. As expected, the models all have generally similar DIC values for the first two settings, which are just progressing and stable representations of PLR. As the data generating settings increase in model complexity, the DIC value corresponding to Spatial CP separates from the non-spatial CP models and PLR. 
The effective number of parameters, $p_D$, is similar for all models in Setting 1, but is generally smaller for Spatial CP and Non-spatial CP (L) for the remaining settings. It is encouraging that Spatial CP has a smaller $p_D$ in the more complex settings, indicating that the increased model structure maintains parsimony even in the presence of complex data.

\setlength{\tabcolsep}{5pt}
\begin{table}[t]
\centering
\caption{Model fit diagnostics for the simulation study. The diagnostics included are deviance information criterion (DIC) and effective number of parameters ($p_D$) and are presented for each of the models introduced in Section \ref{sec:mc}. The data generating settings for the simulation increase in model complexity and are as follows, 1) progressing, 2) stable, 3) non-spatial CP, 4) spatial CP with no covariance, and 5) full spatial CP model. Each reported estimate is based on 250 simulated datasets.\label{tab:sim1}} 
\begin{tabular}{llrrrrr}
  \hline 
& & \multicolumn{5}{c}{Simulation Setting} \\ \cline{3-7} 
Metric & Model & 1 & 2 & 3 & 4 & 5 \\ 
  \hline
DIC & PLR & -466.33 & -409.93 & -360.16 & -401.16 & -276.76 \\ 
   & Non-spatial CP (D) & -457.89 & -434.85 & -426.27 & -455.50 & -293.29 \\ 
   & Non-spatial CP (C) & -412.21 & -442.09 & -414.03 & -448.02 & -293.54 \\ 
   & Non-spatial CP (L) & -473.37 & -476.06 & -446.44 & -446.77 & -318.36 \\ 
   & Spatial CP & -479.61 & -455.21 & -457.57 & -468.10 & -351.22 \\ \hline
$p_D$ & PLR & 118.23 & 118.39 & 118.59 & 121.08 & 118.72 \\ 
   & Non-spatial CP (D) & 111.19 & 118.51 & 117.44 & 134.85 & 125.36 \\ 
   & Non-spatial CP (C) & 133.70 & 105.29 & 120.37 & 135.56 & 115.00 \\ 
   & Non-spatial CP (L) & 109.34 & 69.65 & 87.82 & 78.27 & 77.70 \\ 
   & Spatial CP & 118.80 & 92.41 & 110.22 & 94.71 & 90.56 \\ 
   \hline
\end{tabular}
\end{table}

Next, we studied the capability of each model to predict future VF sensitivities. Prediction ability was assessed for each model by predicting the simulated data for the final 7 visits excluded from the modeling. In Table \ref{tab:sim2}, we present results for two of these visits. For Setting 1, MSPE shows similar values for most models within each time point with the exception of Spatial CP which is superior to the competing methods. This is surprising, since this data generating setting is a subset of each of the competing models, but possibly indicates that Spatial CP is preferred even in the absence of clear spatial structure. In Setting 2, all of the models have comparable performance, while in Settings 3 and 4, Spatial CP generally shows small improvements. Finally, in the most complex setting, Spatial CP has superior prediction ability at both future visits.

\setlength{\tabcolsep}{5pt}
\begin{table}[t]
\centering
\caption{Prediction diagnostics for estimation of the visual field (VF) visits at future times 0.75 and 1. Mean squared prediction error is presented for each of the models introduced in Section \ref{sec:mc}. The data generating settings for the simulation increase in model complexity and are as follows, 1) progressing, 2) stable, 3) non-spatial CP, 4) spatial CP with no covariance, and 5) full spatial CP model. Each reported estimate is based on 250 simulated datasets. \label{tab:sim2}} 
\begin{tabular}{llrrrrr}
  \hline 
& & \multicolumn{5}{c}{Simulation Setting} \\ \cline{3-7} 
Time & Model & 1 & 2 & 3 & 4 & 5 \\ \hline
0.75 &   PLR & 215.14 & 577.70 & 450.40 & 358.81 & 531.48 \\ 
&   Non-spatial CP (D) & 213.56 & 572.25 & 447.12 & 364.53 & 532.32 \\ 
&   Non-spatial CP (C) & 215.35 & 577.26 & 451.93 & 362.37 & 531.95 \\ 
&   Non-spatial CP (L) & 213.62 & 577.54 & 451.17 & 359.04 & 530.65 \\  
&   Spatial CP & 187.63 & 577.25 & 439.73 & 349.58 & 524.76 \\  \hline
1 &  PLR & 131.81 & 573.39 & 451.89 & 248.02 & 541.26 \\ 
&   Non-spatial CP (D) & 129.24 & 558.73 & 443.05 & 258.77 & 539.04 \\ 
&    Non-spatial CP (C) & 132.91 & 573.20 & 455.12 & 254.08 & 542.85 \\ 
&    Non-spatial CP (L) & 131.47 & 574.85 & 452.77 & 247.87 & 541.05 \\ 
&    Spatial CP & 109.62 & 575.05 & 443.39 & 239.76 & 527.46 \\ \hline
\end{tabular}
\end{table}

Finally, we investigate the ability of each model to estimate both the observed, $\theta(\mathbf{s})$, and latent, $\eta(\mathbf{s})$, CPs. In Table \ref{tab:sim3}, we present bias, MSE, and EC for models that are capable of estimating either of these CPs. Note that Non-spatial CP (D) was not included in the observed CP section since it is estimating a different, discrete, CP. The results for estimating the observed CP demonstrate the importance of the latent CP process for properly modeling the true CP. The most important finding in these results is that Non-spatial CP (C) is incapable of estimating the observed CP, with inferior bias, MSE, and EC. In particular, the EC is estimated to be zero for Settings 1-2. This is clearly due to the lack of a latent CP specification, as the two models with this feature have improved EC. 
Finally, it is important to note that Spatial CP has superior bias and MSE in estimating the observed CPs, indicating that in addition to the latent CPs, spatial structure is also crucial.

\setlength{\tabcolsep}{0.5pt}
\begin{table}[t]
\centering
\caption{Summary diagnostics for estimation of the observed, $\theta(\mathbf{s})$, and latent, $\eta(\mathbf{s})$, change points (CPs). The diagnostics included are bias, mean squared error (MSE)), and empirical coverage (EC) (95\% nominal) and are presented for each of the models introduced in Section \ref{sec:mc} that are capable of estimating either the observed or latent CP. The data generating settings for the simulation increase in model complexity and are as follows, 1) progressing, 2) stable, 3) non-spatial CP, 4) spatial CP with no covariance, and 5) full spatial CP model. Each reported estimate is based on 250 simulated datasets. \label{tab:sim3}}
\begin{tabular}{lllrrrrr}
  \hline 
& & & \multicolumn{5}{c}{Simulation Setting} \\ \cline{4-8} 
Estimand & Metric & Model & 1 & 2 & 3 & 4 & 5 \\ \hline
$\theta(\mathbf{s})$ & & Non-spatial CP (C) & -0.11 & 0.19 & 0.07 & 0.07 & 0.09 \\ 
&   Bias & Non-spatial CP (L) & -0.00 & 0.06 & 0.01 & 0.04 & 0.01 \\ 
&    & Spatial CP & -0.00 & 0.00 & 0.00 & 0.01 & 0.00 \\ \cline{2-8}
&    & Non-spatial CP (C) & 0.02 & 0.07 & 0.05 & 0.03 & 0.05 \\ 
& MSE   & Non-spatial CP (L) & 0.00 & 0.04 & 0.03 & 0.04 & 0.04 \\ 
&    & Spatial CP & 0.00 & 0.00 & 0.00 & 0.01 & 0.00 \\ \cline{2-8}
& \multirow{3}{*}{EC}   & Non-spatial CP (C) & 0.00 & 0.00 & 0.16 & 0.58 & 0.16 \\ 
&    & Non-spatial CP (L) & 1.00 & 1.00 & 0.98 & 0.90 & 0.98 \\ 
&    & Spatial CP & 1.00 & 1.00 & 0.98 & 0.97 & 0.99 \\ \hline
$\eta(\mathbf{s})$& \multirow{2}{*}{Bias} & Non-spatial CP (L) & 16.00 & -17.69 & -3.82 & -11.18 & -18.99 \\ 
&    & Spatial CP & -1.59 & 1.01 & -0.00 & -0.01 & 0.01 \\ \cline{2-8}
& \multirow{2}{*}{MSE} & Non-spatial CP (L) & 4088.01 & 5399.53 & 4615.01 & 4453.89 & 5531.15 \\ 
&    & Spatial CP & 33.33 & 26.85 & 1.90 & 0.04 & 0.56 \\ \cline{2-8}
& \multirow{2}{*}{EC} & Non-spatial CP (L) & 0.99 & 0.99 & 0.90 & 0.84 & 0.96 \\ 
&    & Spatial CP & 0.85 & 0.84 & 0.90 & 0.91 & 0.95 \\ \hline   
\end{tabular}
\end{table}

In the simulation study for estimation of the latent CPs, only Spatial CP and Non-spatial CP (L) are compared (i.e., the only models with latent CPs). It is immediately clear that the inclusion of spatial structure is critical for estimating the latent CPs. In all simulation settings, the bias and MSE are improved for Spatial CP. The bias and MSE are generally larger in the Settings 1-2 because there are no observed CPs in the generated data. 
From Table \ref{tab:sim3} it is clear that Spatial CP is needed for accurately estimating both the latent and observed CPs. 

Finally, an additional simulation study is presented in \ref{sec:simlast} to demonstrate the performance of Spatial CP in estimating the observed CP across various points of follow-up. In Figure \ref{fig:simlast} and Table \ref{tab:simlast}, the results illustrate the importance of Spatial CP, especially in models with CPs near the bounds of follow-up.


\section{Discussion \label{sec:disc3}}

In this paper, we presented a unified framework that combines multiple vital aspects of glaucoma management; i) prediction of future VF sensitivities, ii) predicting the timing and spatial location of future vision loss, and iii) making clinical decisions regarding progression. Furthermore, this framework incorporates important anatomical information. Previous statistical work has focused on one or some of these aspects typically, but to our knowledge, no methods have considered all simultaneously; in particular the latent CP process. Although motivated by VF data, the methodology was introduced in a general manner that permits the model to be applied in broad areal data settings where a CP is appropriate. The methodology is built upon theory in the CP literature, utilizing a Bayesian hierarchical modeling framework for making inference. 

We extended previous attempts at modeling CPs across spatial units, that either ignore spatial dependencies \citep{wagner2014modeling} or do not include a CP in the variance process \citep{warren2017spatial}. The inclusion of the variance CP process can add power to detect CPs and was motivated by the known inverse relationship between decreases in VF sensitivity and variability \citep{russell2012relationship}. Finally, we extended the univariate spatial boundary detection framework for allowing a dissimilarity metric to dictate the local neighborhood structure on the VF. The novel multivariate extension allows the anatomy of the optic disc to dictate the neighborhood adjacencies for the mean, variance, and CP processes. This specification functions jointly and accounts for known dependencies between the separate processes through a cross-covariance. The combination of these model features leads to a superior modeling technique for longitudinal series of VFs. 

The results from applying our method to VF data (Section \ref{sec:glaucoma3}) demonstrated the benefit of our methodology in both prediction of future VF sensitivities and model fit. Spatial CP was compared to non-spatial CP models and PLR and was shown to have superior DIC and MSPE (Table \ref{tab:diagpred}); this was later confirmed in simulation (Tables \ref{tab:sim1} and \ref{tab:sim2}). These results indicate the promise of Spatial CP clinically, as predicting future VF sensitivities is incredibly important for managing progression \citep{crabb1997improving}. These improvements can be attributed to the inclusion of a spatial dependency structure and the flexible trajectories produced by the CPs across time. The need for spatiotemporal treatment of VF data is demonstrated here by a reduction in prediction errors and observed clinically meaningful results.

In addition to an increase in prediction precision, the CP framework allows for prediction of the timing and spatial location of future vision loss. This information is accessible due to both the modeling of spatial dependencies and the latent specification of the CP. The benefits of the latent CP are that it provides a biological interpretation of true disease progression and informs about VF sensitivity outside of the observed follow-up period. The latent CP is modeled using the novel multivariate spatial process, which allows the anatomy of the optic disc to dictate the local neighborhood structure. As such, the patterns of future vision loss based on the latent CP are anatomically driven. This can be seen in the estimated probabilities of a CP for an example patient (right panel of Figure \ref{fig:vfoutput} and Video \ref{vid:gif}), which display a clear differentiation in the inferior and superior sectors as defined by \citet{garway2000mapping}. The practical utility of this latent CP specification is rooted in the biological motivation, but also in that it provides an alternate method for predicting severity across the VF. 

The latent CP clearly represents an improvement over standard CP specifications that limit the range of the CP to the observed follow-up period. Furthermore, it allows us to predict a unobserved CPs. This was exemplified with an example patient with many observed CPs near the end of follow-up. For this patient, we showed that when removing the last two visits (and thus truncating these CPs), Spatial CP was still capable of detecting the presence of an imminent CP. This served as a proof of practice for predicting unobserved CPs and was facilitated by the spatial correlation between VF locations, as Non-spatial CP (L) was incapable of detecting these truncated CPs (Figure \ref{fig:extra}). This was further confirmed in simulation, as Spatial CP estimated the latent CP with lower bias and MSE (Table \ref{tab:sim3}).

We defined a diagnostic metric as the maximum probability across the VF of a CP occurring in the observed follow-up period and showed that it is a significant predictor of clinically determined glaucoma progression (Table \ref{tab:diagpred}). This contribution is important clinically, since a detectable change in a longitudinal series of VFs can be highly suggestive of progression, even if standard VF metrics may appear well within normal limits \citep{artes2005longitudinal}. This finding supports the hypothesis that a CP can intrinsically parameterize the timing of progression. Furthermore, it confirms the importance of Spatial CP for characterizing structural change on the VF that can be used for improved clinical decision making.

This paper introduced a framework for assessing glaucoma progression from numerous angles and demonstrates the clinical utility of CPs acting as a proxy for disease progression. To validate these findings it will be important to compare the diagnostic efficiency of the CPs to other established metrics in the glaucoma progression literature, in addition to the ones presented here. Furthermore, the estimation of progression can be improved when combining functional (e.g., VF) and structural (e.g., optic disc) testing techniques \citep{nicolela2003visual}. As such, a synthesis of these two data sources could provide improvement in estimating the CPs. This would provide stronger clinical evidence of progression as damage to the optic disc could be corroborated with the VF and vice-versa \citep{artes2005longitudinal}.  Finally, this work opens up numerous avenues for future statistical research including an extension to consider multiple CPs within a single spatial location and the generalization of the separable multivariate spatial process to a more flexible structure with varying dissimilarity metric parameters, $\alpha_p$ (e.g., linear model of coregionalization). 


\section*{Acknowledgements}
This work was partially supported by the National Center for Advancing Translational Science (JLW; UL1 TR001863, KL2 TR001862) and the Research to Prevent Blindness (Department of Ophthalmology, University of North Carolina-Chapel Hill). The authors thank Brigid D.\ Betz-Stablein (School of Medical Sciences, University of New South Wales), William H.\ Morgan (Lions Eye Institute, University of Western Australia), Philip H.\ House (Lions Eye Institute, University of Western Australia), and Martin L.\ Hazelton (Institute of Fundamental Sciences, Massey University) for providing the dataset from their original study for use in this analysis.


\appendix


\section{Implementation of the model}
\label{sec:mcmcappndx}

For a full introduction of the model, see Sections \ref{sec:methods3} and \ref{sec:glaucoma3} of the main text. Recall that to induce the Tobit model, we define a latent process through the standard Tobit link, $Y_t^*\left(\mathbf{s}\right) = \max\left\{0,Y_t\left(\mathbf{s}\right)\right\}$. As mentioned in the main text, to appropriately model the course of glaucoma progression, we will modify the mean and variance structures to reflect the glaucoma follow-up. The mean and variance processes are defined as follows,
\begin{equation} \notag
\begin{split}
\mu_t\left(\mathbf{s}\right)=\left\{ \begin{array}{ll}
        {\beta}_0\left(\mathbf{s}\right) & \mbox{if $0 \leq x_t \leq \theta\left(\mathbf{s}_i\right)$},\\
        {\beta}_0\left(\mathbf{s}\right) + {\beta}_1\left(\mathbf{s}\right)\left\{x_t-\theta\left(\mathbf{s}\right)\right\} & \mbox{if $\theta\left(\mathbf{s}\right) \leq x_t \leq x_{\nu}$},\end{array} \right.
\end{split}
\end{equation}
and
\begin{equation} \notag
\begin{split}
\log\{{\sigma_t}\left(\mathbf{s}\right)\}=\left\{ \begin{array}{ll}
        {\lambda}_0\left(\mathbf{s}\right) & \mbox{if $0 \leq x_t \leq \theta\left(\mathbf{s}\right)$},\\
        {\lambda}_0\left(\mathbf{s}\right) + {\lambda}_1\left(\mathbf{s}\right)\left\{x_t-\theta\left(\mathbf{s}\right)\right\} & \mbox{if $\theta\left(\mathbf{s}\right) \leq x_t \leq x_{\nu}$}.\end{array} \right.
\end{split}
\end{equation}
To obtain this form from the specification detailed in Section \ref{sec:spcpmodel} of the main text, set $\beta_1(\mathbf{s}) = 0$ and then recast $\beta_2(\mathbf{s})$ as $\beta_1(\mathbf{s})$. The same is done with the variance parameters. Finally, we force the CPs from the mean and variance to be equal.

As mentioned in Section \ref{sec:glaucoma3}, we draw samples from the joint posterior using MCMC techniques \citep{metropolis1953equation,geman1984stochastic,gelfand1990sampling}. The Tobit model specification is particularly amenable to MCMC, since the latent process has a closed form full conditional distribution. To guarantee that Metropolis acceptance rates are within an acceptable range, pilot adaptation is used \citep{banerjee2003hierarchical}. To improve MCMC mixing, the VF data were scaled by 10, time was parameterized in years, and the dissimilarity metric was scaled by 100. The MCMC sampler is implemented in the R package \texttt{spCP}.


\subsection{Change point likelihood}
\label{sec:cp}

\noindent The change point likelihood can be written for the latent Tobit process, 
\begin{align*}
&f\left\{Y_t\left(\mathbf{s}_i\right)\Big|{\beta}_0\left(\mathbf{s}_i\right),{\beta}_1\left(\mathbf{s}_i\right),{\lambda}_0\left(\mathbf{s}_i\right),{\lambda}_1\left(\mathbf{s}_i\right),\theta\left(\mathbf{s}_i\right)\right\}\\
&= \prod_{t=1}^{\nu} \prod_{i=1}^{m}\left[\text{N}\left({\beta}_0\left(\mathbf{s}_i\right), e^{2{\lambda}_0\left(\mathbf{s}_i\right)}\right)\right]^{1(0 \leq x_t \leq \theta\left(\mathbf{s}_i\right))}  \\
&\times \left[\text{N}\left({\beta}_0\left(\mathbf{s}_i\right) + {\beta}_1\left(\mathbf{s}_i\right) \{x_t - \theta\left(\mathbf{s}_i\right)\}, e^{2[{\lambda}_0\left(\mathbf{s}_i\right) + {\lambda}_1\left(\mathbf{s}_i\right)\{x_t - \theta(\mathbf{s}_i)\} ]}\right)\right]^{1(\theta\left(\mathbf{s}_i\right) \leq x_t \leq x_{\nu})}.
\end{align*}
This likelihood is informative, however for computational purposes we will write the likelihood in matrix form. The latent responses can be written more concisely using an indicator variable. Define the event $\Upsilon_t\left(\mathbf{s}_i\right) = 1\{\theta\left(\mathbf{s}_i\right) \leq x_t \leq x_{\nu}\}$. Then,
\begin{align*}
Y_t\left(\mathbf{s}_i\right) &= {\beta}_0\left(\mathbf{s}_i\right) + {\beta}_1\left(\mathbf{s}_i\right) \{x_t - \theta\left(\mathbf{s}_i\right)\} \Upsilon_t\left(\mathbf{s}_i\right) + \epsilon_t\left(\mathbf{s}_i\right)\\
&=x_t\left(\mathbf{s}_i; \theta\right) \boldsymbol{\beta} \left(\mathbf{s}_i\right)+ \epsilon_t\left(\mathbf{s}_i\right),
\end{align*}
where $\boldsymbol{\beta}\left(\mathbf{s}_i\right) = \left\{\beta_0\left(\mathbf{s}_i\right), \beta_1\left(\mathbf{s}_i\right)\right\}^T$ and $x_t\left(\mathbf{s}_i; \theta\right) = \left[1, \{x_t - \theta\left(\mathbf{s}_i\right)\} \Upsilon_t\left(\mathbf{s}_i\right)\right]$. Now, we determine the vector of latent observations at each time point, defining $\mathbf{Y}_t=\left\{Y_t\left(\mathbf{s}_1\right), Y_t\left(\mathbf{s}_2\right), \ldots, Y_t\left(\mathbf{s}_m\right)\right\}^T$,
\begin{align*}
\mathbf{Y}_t&=X_t\left(\theta\right) \boldsymbol{\beta} + \boldsymbol{\epsilon}_t,
\end{align*}
where $\boldsymbol{\epsilon}_t = \left\{\epsilon_t\left(\mathbf{s}_1\right),\epsilon_t\left(\mathbf{s}_2\right), \ldots, \epsilon_t\left(\mathbf{s}_m\right)\right\}^T$, the regression coefficient vector $\boldsymbol{\beta} = \left\{\boldsymbol{\beta} \left(\mathbf{s}_1\right)^T, \boldsymbol{\beta} \left(\mathbf{s}_2\right)^T, \ldots, \boldsymbol{\beta} \left(\mathbf{s}_m\right)^T\right\}^T$ and $X_t\left(\theta\right)$ is an $m \times 2m$ block diagonal matrix with blocks entries $x_t\left(\mathbf{s}_i; \theta\right)$. Finally, we write the full latent process, $\mathbf{Y} = \left\{\mathbf{Y}_1^T,\mathbf{Y}_1^T,\ldots,\mathbf{Y}_{\nu}^T\right\}^T$,
\begin{align*}
\mathbf{Y}&=\mathbf{X}\left(\theta\right) \boldsymbol{\beta} + \boldsymbol{\epsilon},
\end{align*}
where $\boldsymbol{\epsilon} = \left\{\boldsymbol{\epsilon}_1^T, \boldsymbol{\epsilon}_2^T, \ldots, \boldsymbol{\epsilon}_{\nu}^T\right\}^T$ and $\mathbf{X}\left(\theta\right)$ is an $m\nu \times 2m$ matrix stacking the $X_t\left(\theta\right)$ matrices. 

Having written the mean process in matrix form, we turn our attention to the variance process. Define,
\noindent $\boldsymbol{\sigma}_t^2 = \left\{\sigma_t^2\left(\mathbf{s}_1\right),\sigma_t^2\left(\mathbf{s}_2\right),\ldots,\sigma_t^2\left(\mathbf{s}_m\right)\right\}^T$. Then $\boldsymbol{\sigma}^2 = \left\{\left(\boldsymbol{\sigma}_1^2\right)^T,\left(\boldsymbol{\sigma}_2^2\right)^T,\ldots,\left(\boldsymbol{\sigma}_{\nu}^2\right)^T\right\}^T$. The variance process can be written using the same form as the mean process,
\begin{align*}
\log\left\{\boldsymbol{\sigma}\right\} &= \mathbf{X}(\theta)\boldsymbol{\lambda} \implies \boldsymbol{\sigma}^2 = \exp\{2\mathbf{X}(\theta)\boldsymbol{\lambda}\},
\end{align*}
where $\boldsymbol{\lambda} = \left\{\boldsymbol{\lambda} \left(\mathbf{s}_1\right)^T, \boldsymbol{\lambda}\left(\mathbf{s}_2\right)^T, \ldots, \boldsymbol{\lambda}\left(\mathbf{s}_m\right)^T\right\}^T$, with $\boldsymbol{\lambda}(\mathbf{s}_i) = \{\lambda_0(\mathbf{s}_i),\lambda_1(\mathbf{s}_i)\}^T$. Define the full likelihood covariance as $\boldsymbol{\Omega} = \text{Diag}\left(\boldsymbol{\sigma}^2\right)$. Then, we can write the matrix likelihood as follows,
\begin{align*}f\left(\mathbf{Y}|\boldsymbol{\beta}, \boldsymbol{\lambda}, \boldsymbol{\theta}\right) &\sim \text{MVN}\left(\mathbf{X}\left(\theta\right) \boldsymbol{\beta}, \boldsymbol{\Omega}\right)\\
\log\left\{\boldsymbol{\sigma}\right\} &= \mathbf{X}\left(\theta\right) \boldsymbol{\lambda},
\end{align*}
where $\boldsymbol{\theta} = \left\{\theta\left(\mathbf{s}_1\right),\theta\left(\mathbf{s}_2\right),\ldots,\theta\left(\mathbf{s}_m\right)\right\}^T$.

Having written the likelihood in a convenient matrix form, we now write the full data likelihood which will be used to find the full conditionals. Note that the likelihood does not include $\boldsymbol{\theta}$, but rather is parameterized to include the generating parameters,
\begin{align*}
&f\left(\mathbf{Y},\boldsymbol{\beta},\boldsymbol{\lambda},\boldsymbol{\eta},\boldsymbol{\delta},\boldsymbol{\Sigma},\alpha\right) \propto f\left(\mathbf{Y}|\boldsymbol{\beta},\boldsymbol{\lambda},\boldsymbol{\eta}\right) \\
&\hspace{1cm}\times f\left(\left\{\boldsymbol{\beta},\boldsymbol{\lambda},\boldsymbol{\eta}\right\}|\boldsymbol{\delta},\boldsymbol{\Sigma},\alpha\right) f\left(\boldsymbol{\delta}\right) f\left(\boldsymbol{\Sigma}\right) f\left(\alpha\right),
\end{align*}
where $\boldsymbol{\eta} = \{\eta(\mathbf{s}_1),\eta(\mathbf{s}_2),\ldots,\eta(\mathbf{s}_m)\}^T$.


\subsection{Full conditional distributions}
\label{sec:fc}

\begin{enumerate}
\item Gibbs Step for $\boldsymbol{\delta}:$

Note that $(\mathbf{1}_m \otimes \boldsymbol{\delta}) = (\mathbf{1}_m \otimes \mathbf{I}_5)\boldsymbol{\delta} = \mathbf{Z}_{\delta}\boldsymbol{\delta}$. Then, 
\begin{align*}
f\left(\boldsymbol{\delta}|\cdot\right) &\propto f\left(\boldsymbol{\phi}|\boldsymbol{\delta},\boldsymbol{\Sigma},\alpha\right) \times f\left(\boldsymbol{\delta}\right)\\
&\propto \exp\left\{-\frac{1}{2}\left[\left\{\boldsymbol{\phi}-\mathbf{Z}_{\delta}\boldsymbol{\delta}_{\beta}\right\}^T \left(\mathbf{Q}(\alpha) \otimes \boldsymbol{\Sigma}^{-1}\right)\left\{\boldsymbol{\phi}-\mathbf{Z}_{\delta}\boldsymbol{\delta}\right\} + \frac{\boldsymbol{\delta}^T \boldsymbol{\delta}}{\kappa^2} \right] \right\}\\
&\propto \exp\Bigg\{-\frac{1}{2}\Bigg[\boldsymbol{\delta}^T \left\{\mathbf{Z}_{\delta}^T\left(\mathbf{Q}(\alpha) \otimes \boldsymbol{\Sigma}^{-1}\right)\mathbf{Z}_{\delta} + \frac{\mathbf{I}_5}{\kappa^2}\right\}\boldsymbol{\delta}\\
&\hspace{1in}-2\boldsymbol{\delta}^T\left\{\mathbf{Z}_{\delta}^T\boldsymbol{\Omega}^{-1}\boldsymbol{\phi}\right\}  \Bigg] \Bigg\}\\
&\sim \text{MVN}(\mathbb{E}_{\delta},\mathbb{C}_{\delta}),
\end{align*}
where $\mathbb{C}_{\delta} = \left\{\mathbf{Z}_{\delta}^T\left(\mathbf{Q}(\alpha) \otimes \boldsymbol{\Sigma}^{-1}\right)\mathbf{Z}_{\delta} + \frac{\mathbf{I}_5}{\kappa^2}\right\}^{-1}$,  $\mathbb{E}_{\delta} = \mathbb{C}_{\delta} \left\{\mathbf{Z}_{\delta}^T\boldsymbol{\Omega}^{-1}\boldsymbol{\phi}\right\}$.

\item Gibbs Step for $\boldsymbol{\beta}:$

It is not immediately obvious, but with some manipulation we can find a closed form full conditional distribution for  $\boldsymbol{\beta}$. The key is that we can find an analytical form of the distribution $f\left(\boldsymbol{\beta}|\boldsymbol{\lambda},\boldsymbol{\eta},\boldsymbol{\delta},\boldsymbol{\Sigma},\alpha\right)$. Recall that, $\mathbb{C}\left(\boldsymbol{\phi}\right) = \mathbf{Q}\left(\alpha\right)^{-1} \otimes \boldsymbol{\Sigma}$. Define, $k = \{1,2\}$ and $j = \{3,4,5\}$ as the indeces of the components of $\boldsymbol{\beta}$ in the vector of random effects $\boldsymbol{\phi}_i$. The mean can be partitioned as,
\begin{equation} \notag
\mathbf{1}_m \otimes \boldsymbol{\delta}=\left[
  \begin{array}{c}
      \mathbf{1}_m \otimes \boldsymbol{\delta}_{k}\\
      \mathbf{1}_m \otimes \boldsymbol{\delta}_{j}\\
  \end{array} \right].
\end{equation}
The covariance matrix can then be partitioned in block form as follows,
\begin{equation} \notag
\boldsymbol{\Sigma}=\left[
  \begin{array}{cc}
      \boldsymbol{\Sigma}_{kk} & \boldsymbol{\Sigma}_{kj} \\
      \boldsymbol{\Sigma}_{jk} & \boldsymbol{\Sigma}_{jj} \\
  \end{array} \right],
\end{equation}
so that,
\begin{equation}
  \mathbf{Q}\left(\alpha\right)^{-1} \otimes \boldsymbol{\Sigma}=\left[
  \begin{array}{cc}
      \mathbf{Q}\left(\alpha\right)^{-1} \otimes \boldsymbol{\Sigma}_{kk} & \mathbf{Q}\left(\alpha\right)^{-1} \otimes \boldsymbol{\Sigma}_{kj} \\
      \mathbf{Q}\left(\alpha\right)^{-1} \otimes \boldsymbol{\Sigma}_{jk} & \mathbf{Q}\left(\alpha\right)^{-1} \otimes \boldsymbol{\Sigma}_{jj} \notag
  \end{array} \right].
\end{equation}
Define $\boldsymbol{\phi}_{ij}$ as the original $\boldsymbol{\phi}_i$ vector, only keeping the indeces included in $j$ and $\boldsymbol{\phi}^j = \left\{\boldsymbol{\phi}_{1j}^T, \boldsymbol{\phi}_{2j}^T,\ldots \boldsymbol{\phi}_{mj}^T\right\}^T$Using the properties of the multivariate normal distribution we can compute the moments of the distribution $f\left(\boldsymbol{\beta}|\boldsymbol{\lambda},\boldsymbol{\eta},\boldsymbol{\delta},\boldsymbol{\Sigma},\alpha\right) \sim \text{MVN}(\mathbb{E}_{k|j},\mathbb{C}_{k|j})$,
\begin{align*}
\mathbb{E}_{k|j}=\mathbb{E}[\boldsymbol{\beta}|\boldsymbol{\lambda},\boldsymbol{\eta},\boldsymbol{\delta},\boldsymbol{\Sigma},\alpha] &= (\mathbf{1}_m \otimes \boldsymbol{\delta}_k) + \left[\mathbf{Q}\left(\alpha\right)^{-1} \otimes \boldsymbol{\Sigma}_{kj}\right]\\
&\hspace{1cm}\times\left[\mathbf{Q}\left(\alpha\right)^{-1} \otimes \boldsymbol{\Sigma}_{jj}\right]^{-1}(\boldsymbol{\phi}^j - (\mathbf{1}_m \otimes \boldsymbol{\delta}_j)) \\
&= (\mathbf{1}_m \otimes \boldsymbol{\delta}_k) + \left[\mathbf{I}_n \otimes \boldsymbol{\Sigma}_{kj}\boldsymbol{\Sigma}_{jj}^{-1}\right]\\
&\hspace{1cm} \times (\boldsymbol{\phi}^j - (\mathbf{1}_m \otimes \boldsymbol{\delta}_j)),
\end{align*}
and
\begin{align*}
\mathbb{C}_{k|j}=\mathbb{C}(\boldsymbol{\beta}|\boldsymbol{\lambda},\boldsymbol{\eta},\boldsymbol{\delta},\boldsymbol{\Sigma},\alpha) &= \left[\mathbf{Q}\left(\alpha\right)^{-1} \otimes \boldsymbol{\Sigma}_{kk}\right]\\
&\hspace{1cm}- \left[\mathbf{I}_n \otimes \boldsymbol{\Sigma}_{kj}\boldsymbol{\Sigma}_{jj}^{-1}\right] \left[\mathbf{Q}\left(\alpha\right)^{-1} \otimes \boldsymbol{\Sigma}_{jk}\right] \\
&=\left[\mathbf{Q}\left(\alpha\right)^{-1} \otimes  \boldsymbol{\Sigma}_{k|j}\right],
\end{align*}
where $\boldsymbol{\Sigma}_{k|j} = \left(\boldsymbol{\Sigma}_{kk} -\boldsymbol{\Sigma}_{kj}\boldsymbol{\Sigma}_{jj}^{-1}\boldsymbol{\Sigma}_{jk}\right)$. Now, we can derive the full conditional for $\boldsymbol{\beta}$.
\begin{align*}
\begin{split}
f\left(\boldsymbol{\beta}|\cdot\right) &\propto f\left(\mathbf{Y}\Big|\boldsymbol{\beta},\boldsymbol{\lambda},\boldsymbol{\eta}\right) \times f\left(\boldsymbol{\beta}|\boldsymbol{\lambda},\boldsymbol{\eta},\boldsymbol{\delta},\boldsymbol{\Sigma},\alpha\right)\\
&\propto \exp\Bigg\{-\frac{1}{2}\bigg[\left\{\mathbf{Y}-\mathbf{X}\left(\boldsymbol{\theta}\right)\boldsymbol{\beta}\right\}^T \boldsymbol{\Omega}^{-1}\left\{\mathbf{Y}-\mathbf{X}\left(\boldsymbol{\theta}\right)\boldsymbol{\beta}\right\}\\
&\hspace{2.5cm}+ \left(\boldsymbol{\beta} - \mathbb{E}_{k|j}\right)^T\mathbb{C}_{k|j}^{-1}\left( \boldsymbol{\beta} - \mathbb{E}_{k|j}\right) \bigg] \Bigg\}\\
&\propto \exp\bigg\{-\frac{1}{2}\Big[\boldsymbol{\beta}^T \left\{\mathbf{X}\left(\boldsymbol{\theta}\right)^T\boldsymbol{\Omega}^{-1}\mathbf{X}\left(\boldsymbol{\theta}\right) + \mathbb{C}_{k|j}^{-1}\right\}\boldsymbol{\beta}\\
&\hspace{2.5cm}-2\boldsymbol{\beta}^T\left\{\mathbf{X}\left(\boldsymbol{\theta}\right)^T\boldsymbol{\Omega}^{-1}\mathbf{Y} + \mathbb{C}_{k|j}^{-1}\mathbb{E}_{k|j}\right\}  \Big] \bigg\}\\
&\sim \text{MVN}(\mathbb{E}_{\beta},\mathbb{C}_{\beta}),
\end{split}
\end{align*}
where the covariance is $\mathbb{C}_{\beta} = \left\{\mathbf{X}\left(\boldsymbol{\theta}\right)^T\boldsymbol{\Omega}^{-1}\mathbf{X}\left(\boldsymbol{\theta}\right) + \mathbb{C}_{k|j}^{-1}\right\}^{-1}$, and the mean is $\mathbb{E}_{\beta} = \mathbb{C}_{\beta} \left\{\mathbf{X}\left(\boldsymbol{\theta}\right)^T\boldsymbol{\Omega}^{-1}\mathbf{Y} + \mathbb{C}_{k|j}^{-1}\mathbb{E}_{k|j}\right\}$.

\item Metropolis Step for  ${\lambda}_0\left(\mathbf{s}_i\right):$

Each ${\lambda}_0\left(\mathbf{s}_i\right)$, $i = 1,\ldots,m$, is sampled using a Metropolis step given the following quantity that is proportional to the full conditional density, 
\begin{align*}
&f\left\{{\lambda}_0\left(\mathbf{s}_i\right)|\cdot\right\} \propto f\left(\mathbf{Y}\Big| \boldsymbol{\beta},\boldsymbol{\lambda},\boldsymbol{\eta}\right) \times f\left(\boldsymbol{\phi}|\boldsymbol{\delta},\boldsymbol{\Sigma},\alpha\right)\\
&\hspace{0.5cm}\propto \prod_{t=1}^{\nu} \left[\text{N}\left({\beta}_0\left(\mathbf{s}_i\right), e^{2{\lambda}_0\left(\mathbf{s}_i\right)}\right)\right]^{1 - \Upsilon_t\left(\mathbf{s}_i\right)}  \\
&\hspace{1cm}\times \left[\text{N}\left({\beta}_0\left(\mathbf{s}_i\right) + {\beta}_1\left(\mathbf{s}_i\right) \{x_t - \theta\left(\mathbf{s}_i\right)\}, e^{2[{\lambda}_0\left(\mathbf{s}_i\right) + {\lambda}_1\left(\mathbf{s}_i\right)\{x_t - \theta(\mathbf{s}_i)\} ]}\right)\right]^{\Upsilon_t\left(\mathbf{s}_i\right)}\\
&\hspace{1cm}\times f\left(\boldsymbol{\lambda}_0|\{\boldsymbol{\beta},\boldsymbol{\lambda}_1,\boldsymbol{\eta}\},\boldsymbol{\delta},\boldsymbol{\Sigma},\alpha\right),
\end{align*}
where $f\left(\boldsymbol{\lambda}_0|\{\boldsymbol{\beta},\boldsymbol{\lambda}_1,\boldsymbol{\eta}\},\boldsymbol{\delta},\boldsymbol{\Sigma},\alpha\right) \sim \text{MVN}\left(\mathbb{E}_{k|j},\mathbb{C}_{k|j}\right)$, with $k = \{3\}$ and $j = \{1,2,4,5\}$. Here $\boldsymbol{\lambda}_j = \{\lambda_j(\mathbf{s}_1),\lambda_j(\mathbf{s}_2),\ldots,\lambda_j(\mathbf{s}_m)\}^T$, $j = 0,1$.

\item Metropolis Step for ${\lambda}_1\left(\mathbf{s}_i\right):$

Each ${\lambda}_1\left(\mathbf{s}_i\right)$, $i = 1,\ldots,m$, is sampled using a Metropolis step given the following quantity that is proportional to the full conditional density, 
\begin{align*}
&f\left\{{\lambda}_1\left(\mathbf{s}_i\right)|\cdot\right\} \propto f\left(\mathbf{Y}\Big| \boldsymbol{\beta},\boldsymbol{\lambda},\boldsymbol{\eta}\right) \times f\left(\boldsymbol{\phi}|\boldsymbol{\delta},\boldsymbol{\Sigma},\alpha\right)\\
&\hspace{0.5cm}\propto \prod_{t=1}^{\nu} \left[\text{N}\left({\beta}_0\left(\mathbf{s}_i\right), e^{2{\lambda}_0\left(\mathbf{s}_i\right)}\right)\right]^{1-\Upsilon_t\left(\mathbf{s}_i\right)}  \\
&\hspace{1cm}\times \left[\text{N}\left({\beta}_0\left(\mathbf{s}_i\right) + {\beta}_1\left(\mathbf{s}_i\right) \{x_t - \theta\left(\mathbf{s}_i\right)\}, e^{2[{\lambda}_0\left(\mathbf{s}_i\right) + {\lambda}_1\left(\mathbf{s}_i\right)\{x_t - \theta(\mathbf{s}_i)\} ]}\right)\right]^{\Upsilon_t\left(\mathbf{s}_i\right)}\\
&\hspace{1cm}\times f\left(\boldsymbol{\lambda}_1|\{\boldsymbol{\beta},\boldsymbol{\lambda}_0,\boldsymbol{\eta}\},\boldsymbol{\delta},\boldsymbol{\Sigma},\alpha\right),
\end{align*}
where $f\left(\boldsymbol{\lambda}_1|\{\boldsymbol{\beta},\boldsymbol{\lambda}_0,\boldsymbol{\eta}\},\boldsymbol{\delta},\boldsymbol{\Sigma},\alpha\right) \sim \text{MVN}\left(\mathbb{E}_{k|j},\mathbb{C}_{k|j}\right)$, with $k = \{4\}$ and $j = \{1,2,3,5\}$.

\item Metropolis Step for ${\eta}\left(\mathbf{s}_i\right):$

Each ${\eta}\left(\mathbf{s}_i\right)$, $i = 1,\ldots,m$, is sampled using a Metropolis step given the following quantity that is proportional to the full conditional density, 
\begin{align*}
&f\left\{{\eta}\left(\mathbf{s}_i\right)|\cdot\right\} \propto f\left(\mathbf{Y}\Big|\boldsymbol{\beta},\boldsymbol{\lambda},\boldsymbol{\eta}\right) \times f\left(\boldsymbol{\phi}|\boldsymbol{\delta},\boldsymbol{\Sigma},\alpha\right)\\
&\hspace{0.5cm}\propto \prod_{t=1}^{\nu} \left[\text{N}\left({\beta}_0\left(\mathbf{s}_i\right), e^{2{\lambda}_0\left(\mathbf{s}_i\right)}\right)\right]^{1-\Upsilon_t\left(\mathbf{s}_i\right)}  \\
&\hspace{1cm}\times \left[\text{N}\left({\beta}_0\left(\mathbf{s}_i\right) + {\beta}_1\left(\mathbf{s}_i\right) \{x_t - \theta\left(\mathbf{s}_i\right)\}, e^{2[{\lambda}_0\left(\mathbf{s}_i\right) + {\lambda}_1\left(\mathbf{s}_i\right)\{x_t - \theta(\mathbf{s}_i)\} ]}\right)\right]^{\Upsilon_t\left(\mathbf{s}_i\right)}\\
&\hspace{1cm}\times f\left(\boldsymbol{\eta}|\{\boldsymbol{\beta},\boldsymbol{\lambda}\},\boldsymbol{\delta},\boldsymbol{\Sigma},\alpha\right),
\end{align*}
where $f\left(\boldsymbol{\eta}|\{\boldsymbol{\beta},\boldsymbol{\lambda}\},\boldsymbol{\delta},\boldsymbol{\Sigma},\alpha\right) \sim \text{MVN}\left(\mathbb{E}_{k|j},\mathbb{C}_{k|j}\right)$, with $k = \{5\}$ and $j = \{1,2,3,4\}$. 

\item Metropolis Step for $\alpha:$

We transform $\alpha$ to the real line to facilitate sampling. Define a new parameter $\Delta=h(\alpha)=\log\left(\frac{\alpha-a_{\alpha}}{b_{\alpha}-\alpha}\right)$, such that $h^{-1}(\Delta)=(b_{\alpha}\exp\{\Delta\}+a_{\alpha})/(1+\exp\{\Delta\})$ and $\left|\frac{\partial}{\partial \Delta} h^{-1}(\Delta)\right|\propto\exp\{\Delta\}/(1+\exp\{\Delta\})^2$. Now we can sample from the transformed proposal distribution, $\Delta^* \sim \text{N}\left(\Delta,\delta\right)$, where $\delta$ is a tuning parameter. Then we can obtain a proposal of $\alpha$, $\alpha^*=h^{-1}(\Delta^*)$. 
\begin{align*}
f\left(\alpha|\cdot\right) &\propto f\left(\boldsymbol{\phi}|\boldsymbol{\delta},\boldsymbol{\Sigma},\alpha\right) \times f\left(\alpha\right) \times \left|\frac{\partial}{\partial \Delta} h^{-1}(\Delta)\right|.
\end{align*}

\item Gibbs Step for $\boldsymbol{\Sigma}:$

Define $\boldsymbol{\Phi}_{5 \times n} = \{\boldsymbol{\beta}_0,\boldsymbol{\beta}_1,\boldsymbol{\lambda}_0,\boldsymbol{\lambda}_1,\boldsymbol{\eta}\}^T$.
Note that $\boldsymbol{\phi}=\text{vec}(\boldsymbol{\Phi})$. Therefore, using the following properties of the Kronecker product and trace, $(B^T \otimes A)\text{vec}(X)=\text{vec}(AXB)$ and $\text{tr}(AB)=\text{vec}(A)^T\text{vec}(B)$ and defining $\boldsymbol{M} = \boldsymbol{\delta} \mathbf{1}_m^T$, we may rewrite the MCAR prior, $f(\boldsymbol{\phi}|\boldsymbol{\delta},\boldsymbol{\Sigma},\alpha)$, as follows,
\begin{align*}
&\hspace{0.25cm}\propto \exp\left\{-\frac{1}{2}\left(\boldsymbol{\phi}- (\mathbf{1}_m \otimes \boldsymbol{\delta})\right)^T\left(\mathbf{Q}(\alpha) \otimes \boldsymbol{\Sigma}^{-1}\right)\left(\boldsymbol{\phi}- (\mathbf{1}_m \otimes \boldsymbol{\delta})\right)\right\}\\
&\hspace{0.25cm}=\exp\left\{-\frac{1}{2}\left(\text{vec}(\boldsymbol{\Phi}) - \text{vec}(\mathbf{M})\right)^T\left(\mathbf{Q}(\alpha) \otimes \boldsymbol{\Sigma}^{-1}\right)\left(\text{vec}(\boldsymbol{\Phi}) - \text{vec}(\mathbf{M})\right)\right\}\\
&\hspace{0.25cm}=\exp\left\{-\frac{1}{2}\text{vec}(\boldsymbol{\Phi} - \mathbf{M})^T\left(\mathbf{Q}(\alpha) \otimes \boldsymbol{\Sigma}^{-1}\right)\text{vec}(\boldsymbol{\Phi} - \mathbf{M})\right\}\\
&\hspace{0.25cm}=\exp\left\{-\frac{1}{2}\text{vec}(\boldsymbol{\Phi} - \mathbf{M})^T\text{vec}\left(\boldsymbol{\Sigma}^{-1}(\boldsymbol{\Phi} - \mathbf{M})\mathbf{Q}(\alpha)\right)\right\}\\
&\hspace{0.25cm}=\exp\left\{-\frac{1}{2}\text{tr}\left((\boldsymbol{\Phi} - \mathbf{M})^T\boldsymbol{\Sigma}^{-1}(\boldsymbol{\Phi} - \mathbf{M})\mathbf{Q}(\alpha)\right)\right\}\\
&\hspace{0.25cm}=\exp\left\{-\frac{1}{2}\text{tr}\left((\boldsymbol{\Phi} - \mathbf{M})\mathbf{Q}(\alpha)(\boldsymbol{\Phi} - \mathbf{M})^T\boldsymbol{\Sigma}^{-1}\right)\right\}\\
&\hspace{0.25cm}=\exp\left\{-\frac{1}{2}\text{tr}\left(\mathbf{S}_{\Phi,\alpha}\boldsymbol{\Sigma}^{-1}\right)\right\},
\end{align*}
where $\mathbf{S}_{\Phi,\alpha}=(\boldsymbol{\Phi} - \mathbf{M})\mathbf{Q}(\alpha)(\boldsymbol{\Phi}- \mathbf{M})^T$. Now, the full conditional for $\boldsymbol{\Sigma}$ is straight forward,
\begin{align*}
f(\boldsymbol{\Sigma}|\cdot) &\propto f(\boldsymbol{\phi}|\boldsymbol{\delta},\boldsymbol{\Sigma},\alpha) \times f(\boldsymbol{\Sigma})\\
&\propto |\mathbf{Q}(\alpha) \otimes \boldsymbol{\Sigma}^{-1}|^{\frac{1}{2}}\exp\left\{-\frac{1}{2}\text{tr}\left(\mathbf{S}_{\Phi,\alpha}\boldsymbol{\Sigma}^{-1} \right)\right\}\\
&\hspace{1cm}\times|\boldsymbol{\Sigma}|^{\frac{-(\xi+p+1)}{2}} \exp\left\{-\frac{1}{2}\text{tr}(\boldsymbol{\Psi} \boldsymbol{\Sigma}^{-1})\right\}\\
&\propto (|\boldsymbol{\Sigma}^{-1}|^{m} |\mathbf{Q}(\alpha)|^{p})^{\frac{1}{2}}\exp\left\{-\frac{1}{2}\text{tr}\left(\mathbf{S}_{\Phi,\alpha}\boldsymbol{\Sigma}^{-1} \right)\right\}\\
&\hspace{1cm}\times|\boldsymbol{\Sigma}|^{\frac{-(\xi+p+1)}{2}} \exp\left\{-\frac{1}{2}\text{tr}(\boldsymbol{\Psi} \boldsymbol{\Sigma}^{-1})\right\}\\
&\propto |\boldsymbol{\Sigma}^{-1}|^{\frac{m}{2}}\exp\left\{-\frac{1}{2}\text{tr}\left([\mathbf{S}_{\Phi,\alpha}+\boldsymbol{\Psi}]\boldsymbol{\Sigma}^{-1} \right)\right\}|\boldsymbol{\Sigma}|^{\frac{-(\xi+p+1)}{2}} \\
&\propto |\boldsymbol{\Sigma}|^{-\frac{m}{2}}\exp\left\{-\frac{1}{2}\text{tr}\left([\mathbf{S}_{\Phi,\alpha}+\boldsymbol{\Psi}]\boldsymbol{\Sigma}^{-1} \right)\right\}|\boldsymbol{\Sigma}|^{\frac{-(\xi+p+1)}{2}} \\
&\propto |\boldsymbol{\Sigma}|^{\frac{-(m+\xi+p+1)}{2}}\exp\left\{-\frac{1}{2}\text{tr}\left([\mathbf{S}_{\Phi,\alpha}+\boldsymbol{\Psi}]\boldsymbol{\Sigma}^{-1} \right)\right\}\\
&\sim \text{Inverse-Wishart}\left(m+\xi,\mathbf{S}_{\Phi,\alpha}+\boldsymbol{\Psi} \right).
\end{align*}

\item Gibbs Step for $Y_t\left(\mathbf{s}_i\right):$

The latent process $Y_t\left(\mathbf{s}_i\right)$ is sampled from their full conditional distribution of, 
$$f\left\{Y_t\left(\mathbf{s}_i\right)|Y^*_t\left(\mathbf{s}_i\right),\boldsymbol{\Omega}\right\} \propto f\left\{Y_t^*\left(\mathbf{s}_i\right)|Y_t\left(\mathbf{s}_i\right)\right\} \times f\left\{Y_t\left(\mathbf{s}_i\right)|\boldsymbol{\Omega}\right\},$$ where $\boldsymbol{\Omega}$ is a vector of the model parameters. Then, we know that there are two possibilities, $Y_t^*\left(\mathbf{s}_i\right)=0$ or $Y_t^*\left(\mathbf{s}_i\right)=Y_t\left(\mathbf{s}_i\right)$.  If $Y_t^*\left(\mathbf{s}_i\right)=0$, then,
$$f\left\{Y_t^*\left(\mathbf{s}_i\right)|Y_t\left(\mathbf{s}_i\right)\right\}=P\left[Y_t^*\left(\mathbf{s}_i\right)=0|Y_t\left(\mathbf{s}_i\right)\right]=
 \left\{ 
  \begin{array}{l l}
    1 & \quad Y_t\left(\mathbf{s}_i\right) \leq 0, \\
    0 & \quad Y_t\left(\mathbf{s}_i\right) > 0.\\
  \end{array} \right.
$$
Therefore, $f\left\{Y_t\left(\mathbf{s}_i\right)|Y^*\left(\mathbf{s}_i\right),\boldsymbol{\Omega}\right\} \propto 1\left\{Y_t\left(\mathbf{s}_i\right) \leq 0\right\}f\left\{Y_t\left(\mathbf{s}_i\right)|\boldsymbol{\Omega}\right\}$. Furthermore, note that when $Y_t^*\left(\mathbf{s}_i\right)=Y_t\left(\mathbf{s}_i\right)$,
$$f\left\{Y_t\left(\mathbf{s}_i\right)|Y_t^*\left(\mathbf{s}_i\right),\boldsymbol{\Omega}\right\}=f\left\{Y_t\left(\mathbf{s}_i\right)|Y_t\left(\mathbf{s}_i\right),\boldsymbol{\Omega}\right\}=Y_t\left(\mathbf{s}_i\right).$$ Then, we see that the full conditional for the latent variable is,
$$f\left\{Y_t\left(\mathbf{s}_i\right)|Y_t^*\left(\mathbf{s}_i\right),\boldsymbol{\Omega}\right\} \sim
 \left\{ 
  \begin{array}{l l}
    \text{N}^{-}\left(\mu_t\left(\mathbf{s}_i\right),\sigma_t^2\left(\mathbf{s}_i\right)\right) & \text{ } Y_t^*\left(\mathbf{s}_i\right) = 0, \\
    Y_t^*\left(\mathbf{s}_i\right) & \text{ } Y_t^*\left(\mathbf{s}_i\right) = Y_t\left(\mathbf{s}_i\right).\\
  \end{array} \right.
$$
where, $\text{N}^{-}\left(\mu,\sigma^2\right)$ specifies a truncated normal distribution (truncated above by zero).

\item Repeat steps 1-8 until convergence has been achieved and an adequate number of posterior samples have been obtained post-convergence.

\end{enumerate}


\renewcommand{\figurename}{Video}

\section{Prediction of future change points}
\label{sec:gif}

See Video \ref{vid:gif} for a map of VF progression over the course of follow-up and into the future. The print version of this manuscript shows a still frame that matches the right of Figure \ref{fig:vfoutput}. The video can be found on Github: \texttt{https://github.com/berchuck/spCP}.

\setcounter{figure}{0}    


\begin{figure}t
\begin{center}
\includegraphics[scale = 0.65,trim = 0cm 3cm 0cm 0cm, clip=true]{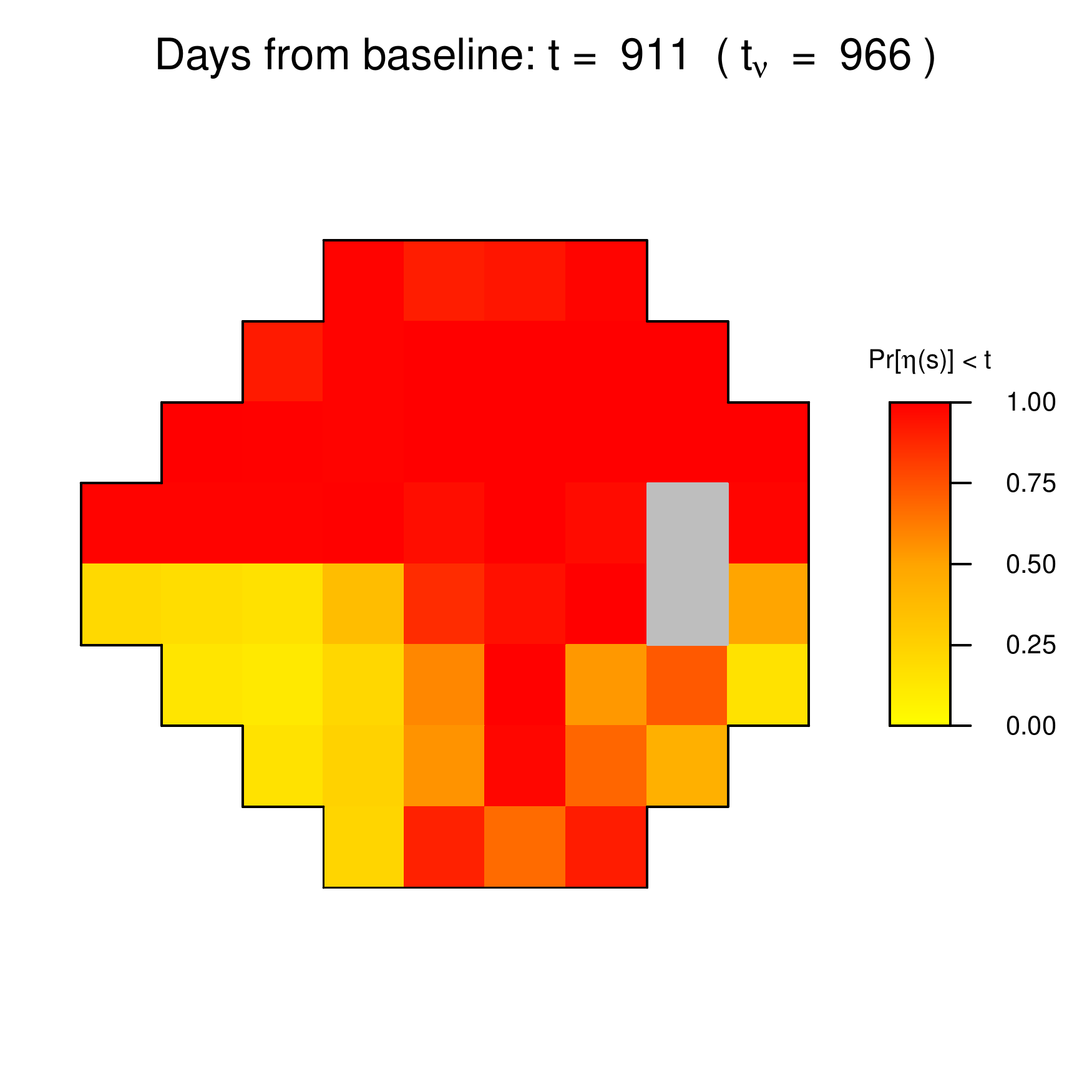}\\
\caption{Clinical video generated from Spatial CP for the same example patient in Figure 1 of the main text. At each location on the VF is the posterior probability of an observed CP presented throughout follow-up and predicted one year into the future. By presenting the CPs in video format, the present and future pattern of progression becomes clear.\label{vid:gif}}
\end{center}
\end{figure}


\section{Estimation of the observed change point}
\label{sec:simlast}

In Section \ref{sec:glaucoma3} of the main text, we established the importance of estimating the observed CPs for identifying the moment of progression. In Section \ref{sec:simstudy3}, we established in simulation the importance of the latent CP specification and spatial structure for estimating both the observed and latent CPs. As a result, it is important to understand the behavior of each modeling feature in estimating CPs, especially at the censored spatial locations. Therefore, we develop a second simulation study that is based on a fixed $\boldsymbol{\phi}$, so that the CPs are fixed across all simulated datasets. The realization of $\boldsymbol{\phi}$ was chosen to produce CPs that occur across all stages of follow-up. To generate the $\boldsymbol{\phi}$, we use the same data generating scheme as Section 6 of the main text, with the same values of $\boldsymbol{\Sigma}$ and $\alpha$, with $\boldsymbol{\delta} = (25, -30, 1, 0.5, 0.5)$. Based on the value of $\boldsymbol{\phi}$, 1000 datasets are simulated. In this simulation, the VF visit times are defined as all 21 visits from Section \ref{sec:simstudy3} of the main text. 

\renewcommand{\figurename}{Figure}

\setcounter{figure}{0}    

\begin{figure}
\begin{center}
\includegraphics[scale=0.7, trim=1.7cm 0.25cm 0cm 0cm, clip=true]{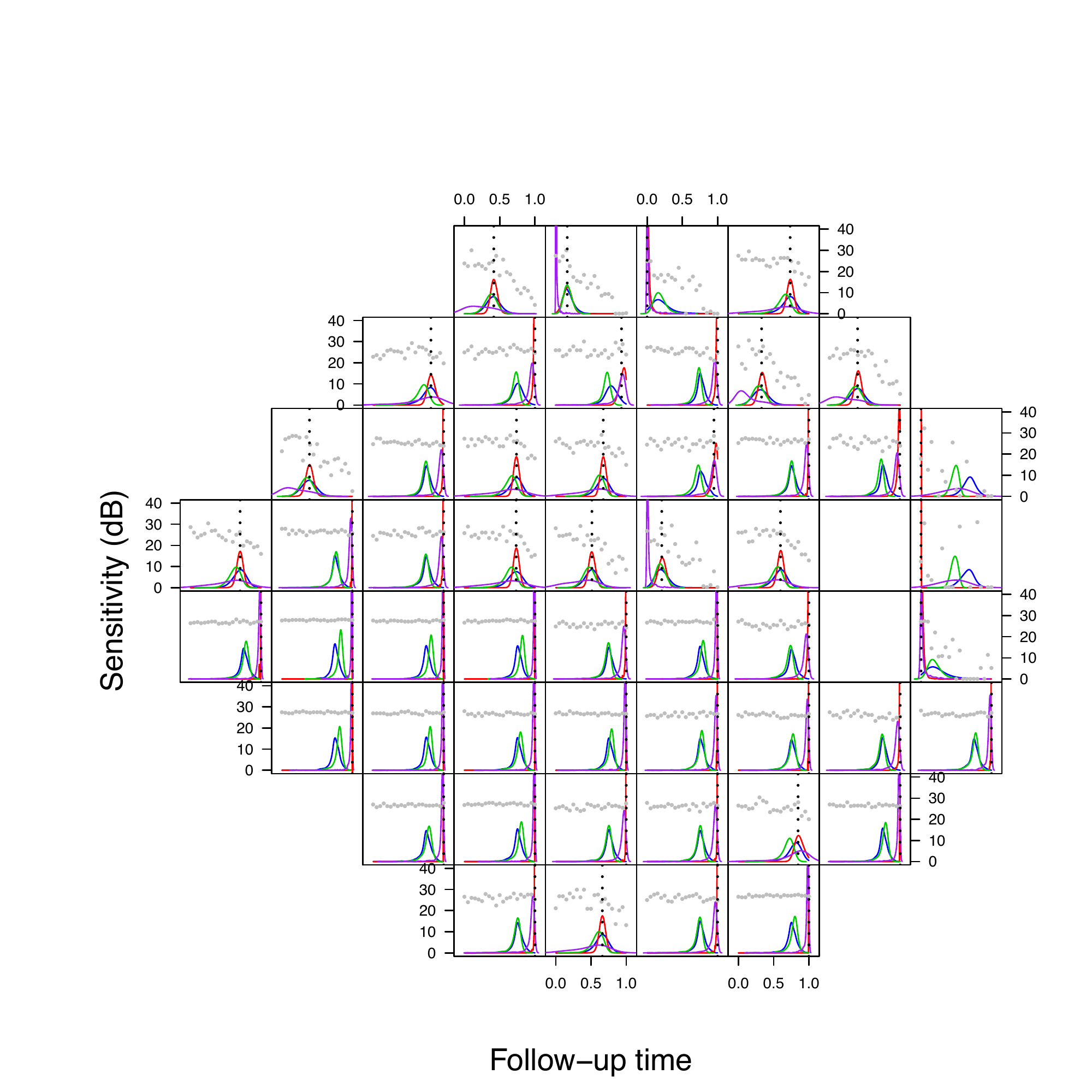}\\
\caption{Simulation study comparing the performance of Spatial CP and non-spatial CP models in estimating the CPs across the VF. The true value of the CP is presented as a dotted black line and the threshold sensitivity values for an example simulated dataset are presented as gray dots. The posterior mean estimates of the CPs across all 1000 datasets for Spatial CP, Non-spatial CP (D), Non-spatial CP (C), and Non-spatial CP (L) are presented as density estimates in red, blue, green, and purple lines, respectively. \label{fig:simlast}}
\end{center}
\end{figure}

Each of the models were fit to the simulated datasets yielding an estimate of the observed CPs at each VF location. In Figure \ref{fig:simlast}, the density estimates of the distribution of estimated observed CPs (posterior means) are presented for each model across the VF (excluding PLR). Furthermore, the true observed CP is presented as a vertical black dotted line and the threshold sensitivity values for an example simulated dataset are presented as gray dots. This figure provides important information regarding the introduced model. First, in VF locations where the true CP is near the middle of the monitoring period (i.e., easier to identify), Spatial CP has minimal bias and has the narrowest bounds (representative of the variability in the posterior mean estimates). In these settings, the other models generally have small bias, however the density estimates are much wider. 

In locations where the true CP is near or beyond the beginning or end of the monitoring period, Spatial CP is clearly superior. The two models that have no latent CP process are incapable of approaching the edges of the follow-up times, with apparent bias. Non-spatial CP (L) provides an improvement, allowing the density estimate to approach the bounds. However, it is important to note that this appears to be a trade-off with wider densities in locations with a true CP in the middle of follow-up. The only model that is capable of producing acceptable CP estimation in the middle and bounds of follow-up is Spatial CP. Interestingly, at certain VF locations at the boundaries, Spatial CP is capable of producing a point mass, meaning all posterior means are the same.

\setcounter{table}{0}    

\setlength{\tabcolsep}{6pt}
\begin{table}[t]
\centering
\caption{Summary diagnostics for estimation of the observed CPs in the simulation detailed in Web Appendix C. The diagnostics included are bias, MSE, and EC (95\% nominal) and are presented for each of the models that are capable of estimating the continuous observed CP. Standard errors (SE) correspond to the variation across all VF locations and simulations. Each reported estimate is based on 1000 simulated datasets.\label{tab:simlast}}
\begin{tabular}{llr}
  \hline
Metric & Model & Estimate (SE) \\ 
  \hline
\multirow{3}{*}{Bias} & Non-spatial CP (C) & 0.147 (0.171) \\    
   & Non-spatial CP (L) & 0.058 (0.123) \\ 
   & Spatial CP & -0.001 (0.008) \\    
   \hline
\multirow{3}{*}{MSE}   & Non-spatial CP (C) & 0.102 (0.065) \\ 
   & Non-spatial CP (L) & 0.083 (0.087) \\ 
 & Spatial CP & 0.002 (0.002) \\    
   \hline
\multirow{3}{*}{EC}   & Non-spatial CP (C) & 0.33 (0.461) \\ 
   & Non-spatial CP (L) & 0.94 (0.146) \\ 
 & Spatial CP & 0.99 (0.018) \\    
   \hline
\end{tabular}
\end{table}

To formalize the results seen in Figure \ref{fig:simlast}, summary diagnostics are presented in Table \ref{tab:simlast}. The results for Non-spatial CP (D) are excluded, since the results are not directly comparable. The results in Table \ref{tab:simlast} confirm the trends observed in Figure \ref{fig:simlast} as the bias and MSE are improved with the introduction of a latent process and then spatial structure. Spatial CP has superior bias and MSE, -0.001 and 0.002, respectively, and has the smallest standard errors, indicating less variability in estimation across the simulated datasets. The EC is only acceptable with the inclusion of the latent CP process. Non-spatial CP (C) has an EC of 0.33, which is clearly a result of all of the true CPs near the bounds. Spatial CP slightly overestimates the nominal coverage, which corresponds to the simulation from the main text and makes sense since the observed CP is a censored parameter. Overall, this additional simulation provides a closer look into the estimation of the observed CP, demonstrating the behavior of each model at various locations over follow-up. These results again illuminate the importance of Spatial CP for estimating the observed CP, especially in models with CPs near the bounds of follow-up, which are commonly the interesting patients.



\bibliographystyle{elsarticle-harv.bst}
\bibliography{References/references.bib}

\end{document}